\documentclass[preprint,12pt,authoryear]{elsarticle}
\usepackage{setspace}

\usepackage{tabularx}
\usepackage{amsmath}
\usepackage[breaklinks=true]{hyperref}

\usepackage{booktabs}
\usepackage{float}
\usepackage{placeins}  
\usepackage{longtable}
\usepackage{multirow}
\usepackage{makecell}

\usepackage{array}
\newcolumntype{P}[1]{>{\centering\arraybackslash}p{#1}}

\usepackage{caption}
\usepackage{amssymb}

\journal{Travel Behaviour and Society}

\begin{document}
\doublespacing

\begin{frontmatter}

\author[aff1]{Daniela Opitz\corref{cor1}}
\ead{[daniela.opitz@usm.cl]}
\cortext[cor1]{Corresponding author}

\author[aff2,aff3]{Eduardo Graells-Garrido}
\author[aff4]{Jacqueline Arriagada}
\author[aff2]{Matilde Rivas}
\author[aff2]{Natalia Meza}

\title{Impact of Shared E-scooter Introduction on Public Transport Demand: A Case Study in Santiago, Chile}

\affiliation[aff1]{organization={Universidad Técnica Federico Santa María},
addressline={Avenida Vicuña Mackenna 3939, San Joaquín},
city={Santiago},
country={Chile}}

\affiliation[aff2]{organization={Department of Computer Science, Universidad de Chile},
addressline={Beauchef 851},
city={Santiago}, 
country={Chile}}

\affiliation[aff3]{organization={National Center for Artificial Intelligence (CENIA)},
addressline={Vicuna Mackenna 4860, Macul},
city={Santiago}, 
country={Chile}}

\affiliation[aff4]{organization={Department of Engineering Technology and Didactics, DTU Engineering Technology},
city={Ballerup}, 
country={Denmark}}

\begin{abstract}
This study examines how the introduction of shared e-scooters affects public transport demand in Santiago, Chile, and whether e-scooters complement or substitute for transit. Using smart-card records and e-scooter GPS traces, we apply a difference-in-differences design with negative binomial regression at the zone level across three urban regions identified by k-means clustering (Central, Intermediate, and Peripheral), controlling explicitly for contemporaneous metro expansion and using randomization inference. In the Intermediate region, which concentrates the service (38 of the 42 treated zones), trips that combine bus and metro within a single journey declined by about 5\% in e-scooter zones ($-4.7\%$ in generation; $-5.7\%$ in attraction). This result is absent in pre-treatment placebo tests, is similar in size with data from a second month, and strengthens with the intensity of e-scooter use, and it is consistent with first/last-mile substitution. By contrast, the change in total public transport demand is not statistically significant (combined bus and metro boardings: $-0.9\%$, 95\% CI: $-3.3$, $+1.6$; alightings: $-1.1\%$, $-3.8$, $+1.6$), in line with the small scale of e-scooter use (0.16\% of boardings in treated zones). At this early stage of adoption, the association between shared e-scooters and public transport is concentrated in journeys that combine bus and metro rather than in total demand.

\end{abstract}


\end{frontmatter}

\section{Introduction}

As cities worldwide invest in sustainable mobility solutions, the effect of shared electric scooters (e-scooters) on public transport is a relevant question for transport planning. In dense urban areas, micromobility (the use of small, lightweight vehicles weighing less than 500 kg \citep{Dediu2019}) can reduce traffic, cut emissions, and improve local economies \citep{tuli2021factors}. However, the impact of e-scooters on transit ridership remains unclear. Previous studies show mixed effects: some classify a substantial share of e-scooter trips as replacing public transport \citep{vinagre2023blind}, while others find e-scooters acting simultaneously as competitors and complements to transit \citep{yan2021spatiotemporal,aarhaug2023scooters}.

Shared e-scooters have emerged as a prominent micromobility option through Mobility as a Service (MaaS) applications \citep{narayanan2023shared}. These services let users locate, reserve and ride scooters via mobile phones, with flexible parking options. Private companies typically operate these services and market them as complementary to public transit, particularly for first/last mile connections \citep{zuniga2022evaluation}. While they could enable modal shifts and complement public transit \citep{latinopoulos2021planning,zuniga2022evaluation}, they also raise concerns about vehicle displacement, unauthorized space use, safety \citep{stigson2021electric,graells2025reducing}, pedestrian conflicts \citep{gossling2020integrating}, and environmental impacts \citep{echeverria2023transitioning}.

Beyond their rapid adoption, e-scooters constitute a mode distinct from shared bicycles: they are free-floating rather than docked, require no physical effort, and are not tied to fixed station locations. Comparative studies show that these attributes translate into spatiotemporal usage patterns and transit relationships that differ from those of station-based bike-share \citep{mckenzie2019spatiotemporal,younes2020comparing}. E-scooters therefore warrant analysis as a distinct intervention rather than as part of an undifferentiated micromobility category, particularly when the question is their effect on public transport demand.

Current knowledge gaps are significant. Most research analyzes e-scooter effects in specific city areas or transit lines, but no study examines how effects vary across different urban structures such as central business districts versus peripheral zones. Additionally, most studies focus on single public transport modes (either bus or metro) rather than evaluating effects on integrated multimodal systems. This represents an important limitation because e-scooters may impact central areas differently than suburban areas, and may affect bus services differently than metro services.

Our study examines three critical research questions: \emph{How does introducing shared e-scooters affect public transport ridership across different urban contexts?} \emph{Are shared e-scooters associated with changes in the journeys that combine bus and metro, where first/last mile effects would operate?} \emph{How do these effects vary by transport mode (bus vs. metro) and urban region?} The answers can help urban planners and transit agencies make informed decisions about e-scooter regulations, infrastructure planning, and integration with existing transit systems.

Santiago (Chile) provides an ideal case study. The city has seen significant transport changes, including reduced transit use and increased motorized transport \citep{graells2023data}, while introducing e-scooters during our study period. Most research examines U.S. and European cities, leaving gaps in understanding how e-scooters affect transit in other contexts. This gap is especially relevant in South America \citep{mitropoulos2023scooter}, where cities face different urban dynamics and transit challenges than well-studied North American and European contexts.

Using a difference-in-differences methodology with comprehensive smart card and GPS data, we advance the field in three ways. First, we show that trips combining bus and metro within a single journey declined by about 5\% in the zones where shared e-scooters were used, while the change in total public transport demand is not statistically significant ($-0.9\%$ in boardings and $-1.1\%$ in alightings in the region that concentrates the service, with confidence intervals that exclude changes larger than about 4\%). The small scale of e-scooter use relative to public transport (0.16\% of boardings in treated zones) helps explain why the substitution and complementarity reported in previous studies need not translate into changes in observed demand. Second, we contribute methodologically by developing an automated approach that uses k-means clustering to group zones into comparable regions, which reduces the arbitrary manual selection of treatment and control zones common in prior quasi-experimental studies, improving the transparency and reproducibility of the comparison; we also estimate effects at the zone level, where treatment is assigned, with randomization inference for regions with few treated zones. Third, we provide evidence by transport mode (bus vs. metro), and for journeys that combine both, in a Latin American context.

The paper proceeds as follows: Section~\ref{sec:litreview} reviews literature on e-scooter impacts. Section~\ref{sec:data} describes our data and study context. Section~\ref{sec:methods} details our methodology, including clustering and regression approaches. Section~\ref{sec:results} presents results, followed by discussion in Section~\ref{sec:discussion}. Section~\ref{sec:conclusions} concludes with implications for urban mobility strategies.

 \section{Literature Review}\label{sec:litreview}

Research on the e-scooter--transit relationship has grown rapidly; its conclusions, however, remain mixed: e-scooters have been reported as substitutes,  competitors, and complements of public transport, with evidence concentrated in North America and Europe \citep{jayawardhena2025shared}. We organize this  literature around three questions: how much e-scooters substitute for versus complement transit, what spatial, temporal, and behavioral factors mediate this relationship, and what causal evidence quasi-experimental designs provide. 
Across all three, significant gaps remain in understanding spatial heterogeneity 
and multimodal effects.

\subsection*{Substitution versus Complementarity}

A first body of work quantifies how often e-scooter trips substitute for or complement transit, and consistently finds both within the same urban area. In Rome, 50\% of e-scooter trips either replaced public transport or served areas with inadequate transit infrastructure \citep{vinagre2023blind}. In Washington, DC, over 70\% of e-scooter trips started and ended within a five-minute walk of transit stops connected by a direct transit line, so the same trips could have been made by transit. The authors caution, however, that these e-scooter trips likely replace walking or represent newly generated travel rather than substitute for transit; only about 10\% of e-scooter trips connected to the Metrorail system as first/last-mile trips \citep{yan2021spatiotemporal}. A study in Oslo similarly found that e-scooters simultaneously compete with and complement transit \citep{aarhaug2023scooters}. The largest multi-city evidence comes from 48{,}301 surveyed e-scooter trips across 20 US cities, where 23.8\% of riders reported connecting to transit before or after the e-scooter trip, while only 3.2\% displaced a transit trip without connecting at all; the complementary e-scooter trips averaged about 1.5 miles, well beyond the typical quarter- to half-mile walking shed, which the authors interpret as e-scooters possibly expanding the catchment area of transit stations \citep{ahmad2025scooters}.

Consistent with a complementary role, proximity and accessibility to transit are repeatedly associated with higher e-scooter use. Greater transit accessibility predicted higher e-scooter ridership in Austin and Minneapolis \citep{bai2020dockless,jiao2020understanding}, and a Louisville study found higher e-scooter trip density in zones combining high walkability with good transit access \citep{hosseinzadeh2021scooters}. A modeling study in Austin further showed that e-scooters serve as a last-mile complement to transit, although the effect was small and confined to the highest-demand zones, such as the university campus and downtown \citep{zuniga2022evaluation}.

\subsection*{Spatial, Temporal, and Behavioral Mediators}

Whether a given trip substitutes for or feeds into transit depends on when, where, and by whom it is taken. Comparative analyses show that e-scooter and station-based bike-share trips follow distinct signatures: e-scooters concentrate in the dense downtown core and support recreational rather than commuting travel, with a midday emphasis instead of the commuting peaks that characterize membership-based bike-share \citep{mckenzie2019spatiotemporal}, and their temporal determinants more closely resemble those of casual than member bike-share users \citep{younes2020comparing}. These patterns also have a within-day structure: in Washington, DC, first-mile e-scooter trips cluster in the morning peak and last-mile trips in the afternoon, so first/last-mile connectivity varies across space and time of day \citep{yan2021spatiotemporal,yin2024shared}. Crucially, this connectivity is hard to measure: the buffer-zone approach commonly used to infer transit-connecting trips from spatial proximity overestimates them substantially, identifying 11.1\% of trips as rail-connecting against only 4.1\% reported by riders themselves \citep{yin2024shared}. On the behavioral side, a stated preference experiment in Seoul found that aggregate mode choices split into 40.6\% shared e-scooter, 39.2\% community feeder bus, and 20.2\% walking, with travelers trading off e-scooters against walking and short transit legs by trip distance and time savings \citep{baek2021electric}. In Norway, about one in five e-scooter trips connects to public transport, and combined use is more likely for trips originating farther from transit \citep{aarhaug2026makes}, consistent with a first/last-mile role that strengthens where access to transit is sparser. The same service can thus act as a substitute or a complement for different users and trip segments.

\subsection*{Quasi-Experimental Evidence}

The studies reviewed so far classify individual trips or associate usage with the built environment; they describe the potential for substitution or complementarity but cannot establish whether transit demand actually changes. A smaller body of work addresses this question with quasi-experimental designs that exploit an intervention, comparing outcomes before and after against a control group, and these designs tend to find small or heterogeneous net effects. For buses, a study in Indianapolis first classified e-scooter trips against the bus network (about 27\% could potentially compete with buses and 29\% complement them) and then used a stop-level difference-in-differences to show that stops exposed to competing trips lost bus ridership after the e-scooter launch \citep{luo2021}. Route-level panel regressions comparing bus ridership before and after e-scooter entry found no significant effect on local routes in Louisville, with suggestive complementary effects on express routes \citep{ziedan2021impacts}, while in Nashville effects differed by inferred trip purpose (utilitarian trips competing and social trips complementing) for a near-zero net effect \citep{ziedan2021complement}. Rail-oriented interventions more consistently show complementarity: a difference-in-differences analysis around new light rail stations in Seattle found significantly increased nearby e-scooter presence after the openings \citep{tyndall2022complementarity}, and an analogous design around a metro line extension in Berlin found e-scooter demand rising near newly opened stations and falling near a closed one \citep{weschke2023scooting}.

\subsection*{Research Gaps}

Current literature exhibits four critical limitations that motivate our study. First, studies examine either spatial patterns or causal effects, but not both within a comprehensive framework that accounts for urban heterogeneity. Second, most research focuses on single transport modes, missing interactions within integrated multimodal networks. Third, research concentrates on North American and European cities, leaving substantial gaps in understanding different urban contexts, particularly in Latin America where transit systems and urban development patterns differ significantly. Fourth, much of the evidence infers first/last-mile connections from the spatial proximity of trip ends to transit stations. Recent work shows that the resulting estimates depend heavily on the buffer distance chosen and vary across areas and times of day \citep{qian2025big}, and that proximity-based counts can substantially overstate the transit connections that riders actually report \citep{yin2024shared}; studies that observe intermodal transfers directly remain scarce.

Our study addresses these gaps by combining quasi-experimental methodology with systematic spatial analysis across an integrated multimodal network in Santiago, Chile. Unlike previous work that manually selects treatment and control zones, we employ automated clustering to enhance methodological rigor while examining how both bus and metro respond across different urban contexts within a single analytical framework, and we identify genuine bus--metro transfers directly from smart-card records rather than inferring multimodal use from spatial proximity.

\section{Data Description}
\label{sec:data}

\subsection*{Study Area}

\begin{figure}
  \centering
  \includegraphics[width=0.95\linewidth]{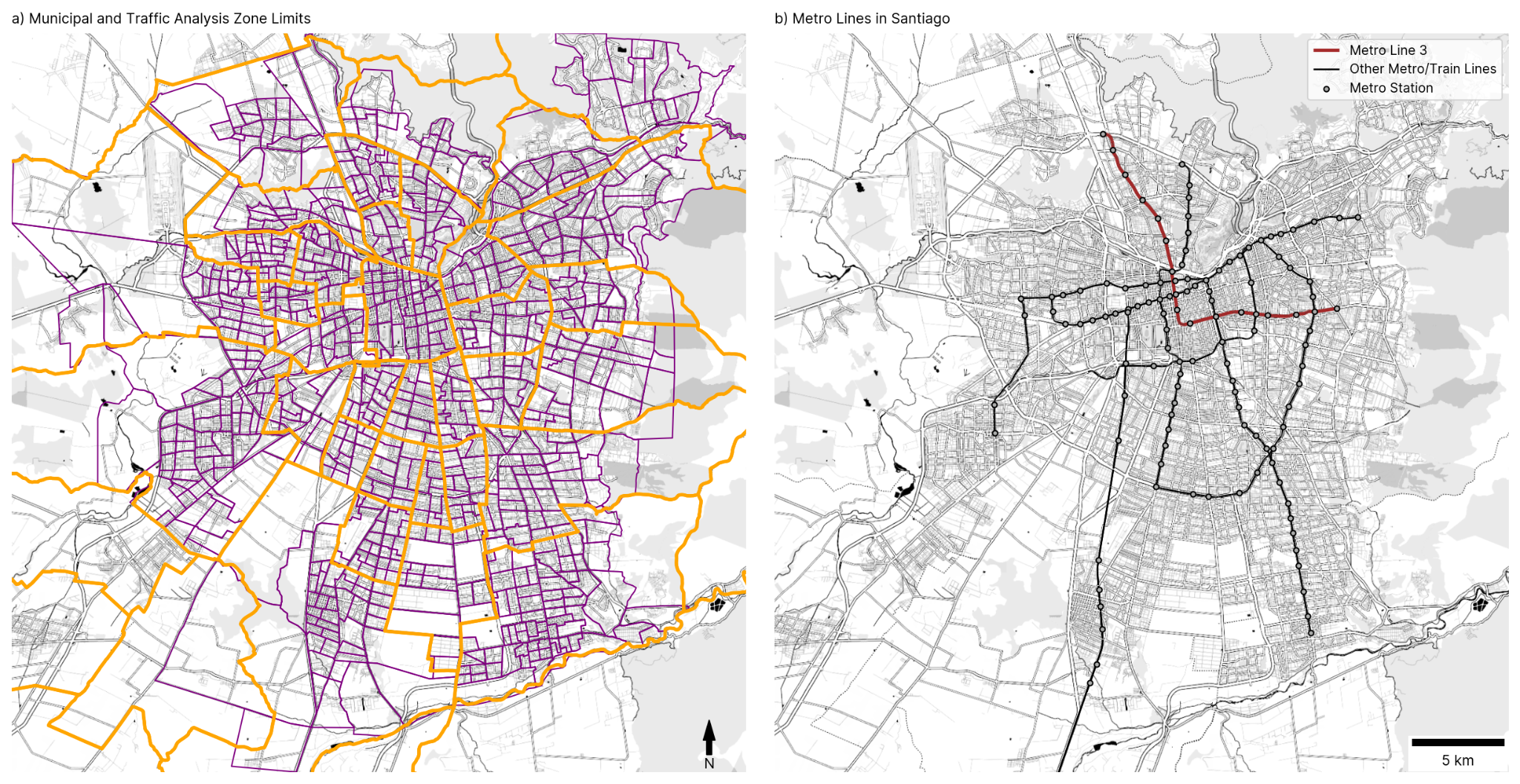}
  \caption{Area under analysis. Left (a): Administrative boundaries of municipalities (orange) and of Traffic Analysis Zones (purple). Right (b): Metro and train lines operating in the city. The red line (Metro Line 3) was inaugurated in January 2019. OpenStreetMap data and Map tiles © CARTO are used under a Creative Commons License.}
  \label{fig:zoning}
\end{figure}

This study examines how e-scooters affect public transport demand in Santiago, the capital and largest city of Chile. As of the 2024 Census \citep{INE2025censo}, Santiago had a population of nearly 8 million across 867.75 square kilometers, comprising over 40 independent municipalities, with 35 in the urban core and the rest on the periphery (see Fig.~\ref{fig:zoning} a).

Our analysis uses geographical zones defined by the Metropolitan Public Transport Directory (DTPM, by its Spanish acronym), comprising 803 Traffic Analysis Zones (TAZs) averaging 1.23 squared km each (see Fig.~\ref{fig:zoning} b).

\subsection*{Data Sources and Study Period}

We analyze two primary data sources across specific time periods. First, we examined \emph{bip!} smart card (\emph{bip!} is the contactless fare card used across public transport in Santiago) records from the working days of the weeks of April 9--13, 2018 and April 8--12, 2019 (Monday through Friday), two equivalent calendar weeks free of national holidays that capture typical urban activity. The 2019 period includes the newly opened Metro Line 3 (inaugurated January 22, 2019). Second, we analyzed GPS traces from shared e-scooters, which operated in Santiago between October 10th, 2018 and April 15th, 2019, covering the introduction and establishment of these services. For the analysis, e-scooter activity is taken from the same week as the smart card records (April 8--14, 2019), so both data sources describe the same period. The dataset includes dockless e-scooters that users could park freely within service areas.

\subsection*{Public Transport Data}

The public transport system of Santiago integrates metro, urban buses, and rail with a nearly flat fare structure. The system allows up to three transfers within two hours using the \emph{bip!} smart card (the only accepted payment method). The card records trip origins but not destinations, as validation occurs only when boarding buses or entering metro stations. Beyond public transport, the city offers taxi services, ride-hailing, and bike-sharing.

\begin{figure}
  \centering
  \includegraphics[width=0.95\linewidth]{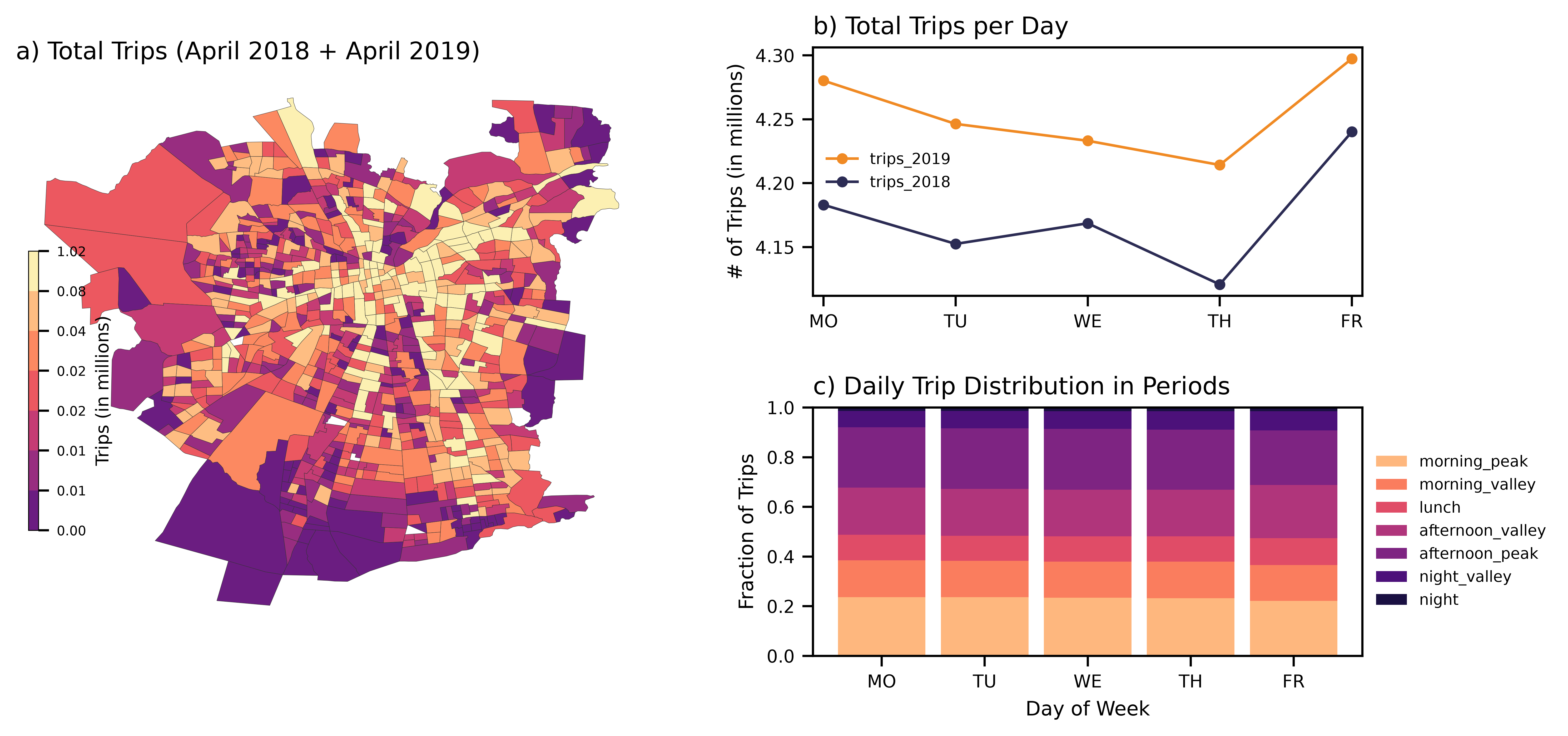}
  \caption{Distributions of Public Transport trips. Left (a): choropleth map of total trips by Traffic Analysis Zone. Top right (b): total trips per day of week, per year. Bottom right (c): relative distribution of trip starting periods for each day of week; time periods are defined in the Supplementary Material (Appendix~A).}
  \label{fig:adatrap}
\end{figure}

The public transport data comes from the Public Transport Data Analysis system (ADATRAP, by its Spanish acronym) of DTPM \citep{Gschwender2016vi}, which consolidates smart card transactions, bus GPS records, and operator information. ADATRAP processes these data using travel pattern analysis and spatial distribution of boarding points \citep{munizaga2012estimation}. We can aggregate trips from bus stops or metro stations into zones (Fig.~\ref{fig:adatrap}(a) shows trip distribution), municipalities, or macro-areas.

\subsection*{Shared E-scooter Data}

\begin{figure}
  \centering
  \includegraphics[width=0.95\linewidth]{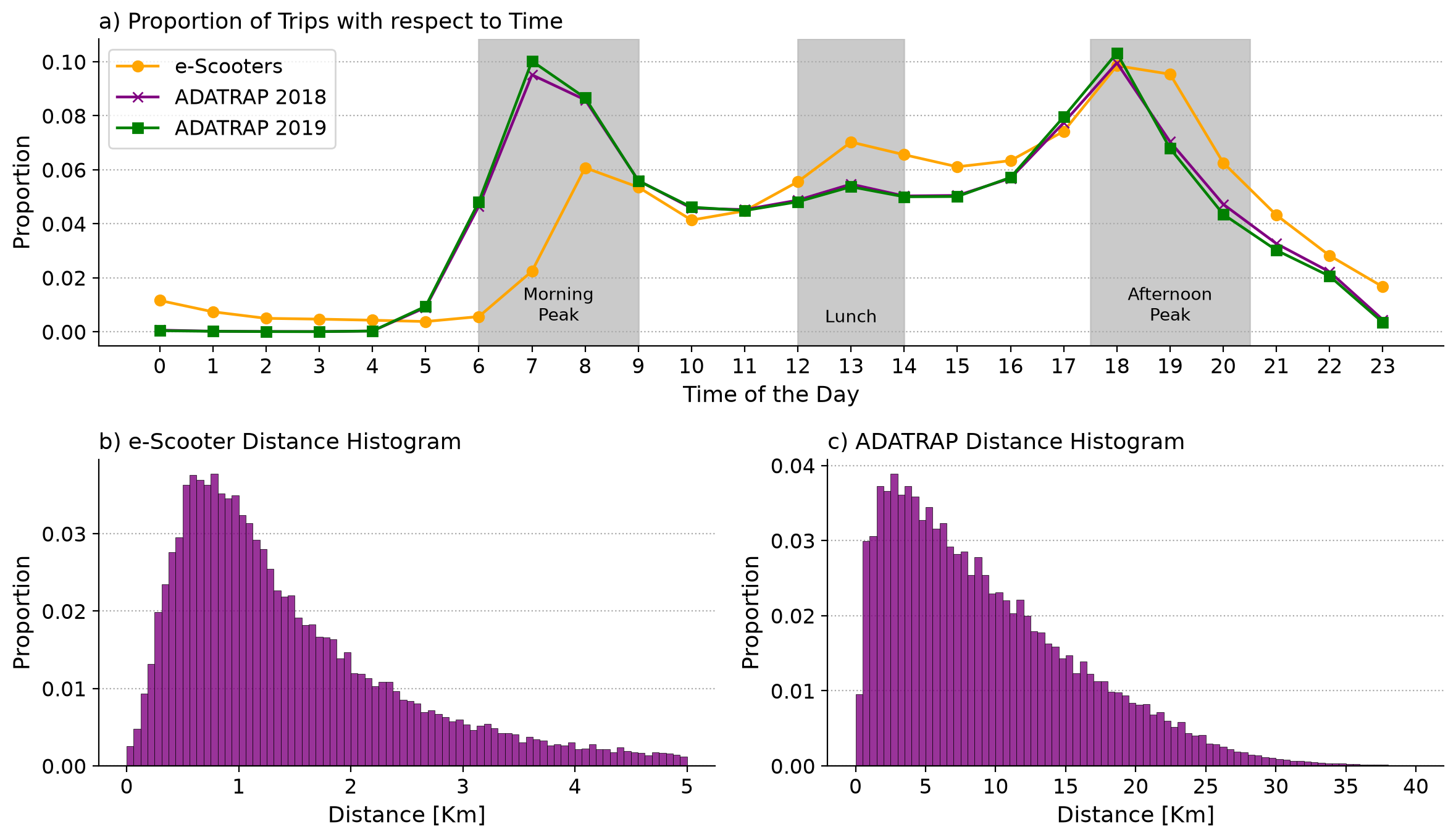}
  \caption{Distributions of trip features from Public Transport and shared e-scooters. Top (a): Proportion of trips with respect to hour of the day, per year. Bottom left (b): Histogram of trip distances in e-scooter trips. Bottom right (c): Histogram of trip distances in Public Transport.}
  \label{fig:tripdistributions}
\end{figure}

\begin{figure}
  \centering
  \includegraphics[width=0.95\linewidth]{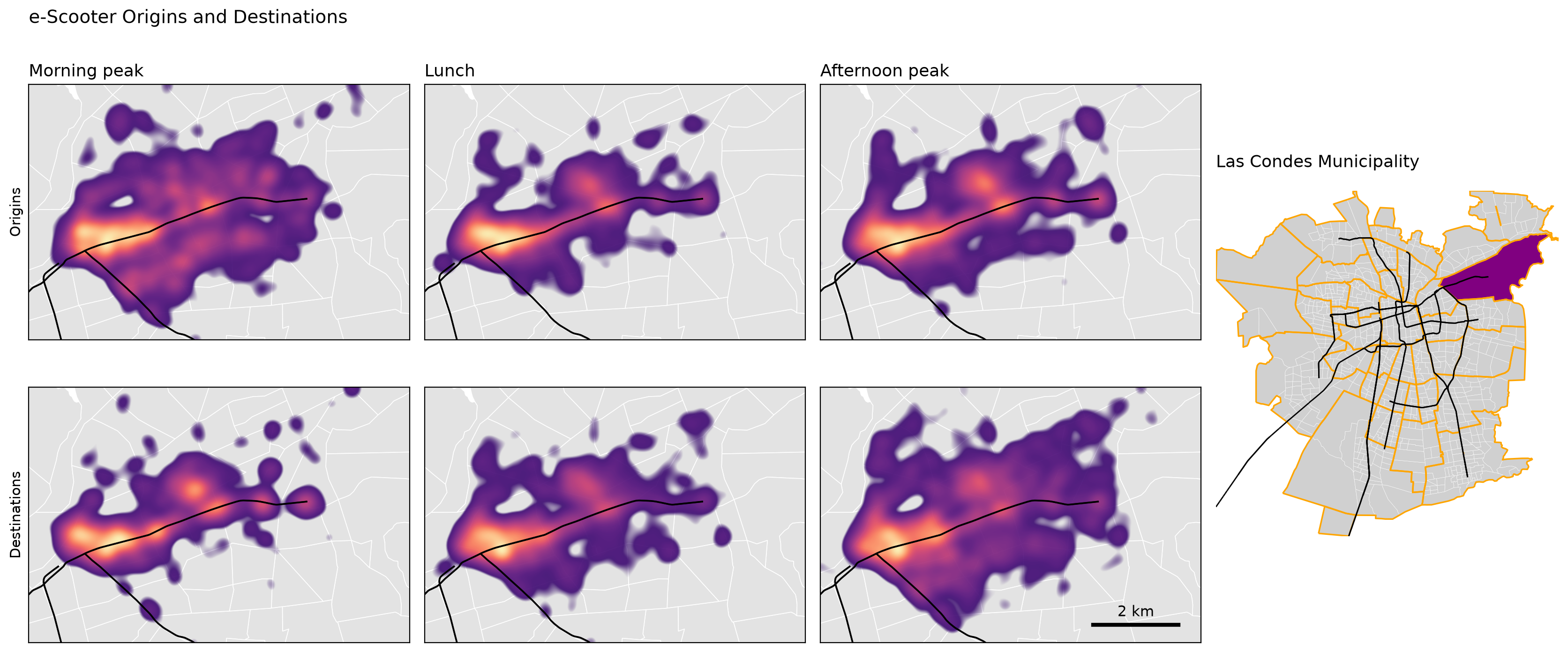}
  \caption{Origins and destinations of shared e-scooter trips. The top row shows trip origins in three periods of the day (morning peak, lunch and afternoon peak), whereas the bottom row shows trip destinations for the same period. The map on the right highlights the area of initial availability of the shared e-scooter service: the municipality of Las Condes.}
  \label{fig:origindestinations}
\end{figure}

We obtained GPS traces of 45K e-scooter trips directly from the operator of the shared e-scooter service (Fig.~\ref{fig:origindestinations}). The service started in the municipality of Las Condes, which concentrates 91\% of the trips, and from February 2019 it extended to parts of the neighboring municipalities of Vitacura and La Reina.  Using a stop-based algorithm \citep{graells2016day}, we identified trips and excluded staff relocation movements. The data reveals distinct patterns from public transport, as shown in Fig.~\ref{fig:tripdistributions}. E-scooter usage shows less pronounced morning peaks from 6:00--9:00, while maintaining similar afternoon peaks between 17:30--20:30. We observed higher lunchtime usage between 12:00--14:00, likely due to the concentration of business areas in Las Condes. Shared e-scooter trips are notably shorter, averaging 1.44 km compared to 9.36 km for public transport, with origins and destinations clustering near metro stations.

Our analysis centers on shared e-scooters because they were the genuinely new shared mode introduced in Santiago during the study period: station-based bike-sharing had operated in the city since 2013, and in Las Condes (where the e-scooter service started) since 2015 \citep{tiznado2021using}. Pre-existing bike-sharing infrastructure was therefore a stable feature of both study periods, whose influence the difference-in-differences design removes along with other time-invariant zone characteristics, whereas the e-scooter roll-out is the intervention of interest. In addition, the operator of the e-scooter service provided georeferenced trip records directly, enabling the zone-level analysis presented here; comparable trip-level data were not available for the contemporaneous bike-sharing systems. A dockless bike-sharing service (Mobike) also operated in Santiago during the study period: it launched in 2018 and operated throughout our 2018--2019 study window \citep{tiznado2021using}. Because it operated in both the pre- and post-treatment periods, it cannot explain a change that appears only after the e-scooter introduction, although the estimates are best read as reflecting the introduction of dockless e-scooters within this broader micromobility context.

\subsection*{Complementary Data}

We complemented these transport datasets with demographic data from the 2017 Chilean Census \citep{censo2017}, including population counts by age and sex, as well as average educational level measured in years of schooling. We used educational level as a proxy for income, an approach supported by research linking education to income distribution \citep{gregorio2002education,jerrim2015income}. Additionally, we used General Transit Feed Specification (GTFS) data from DTPM to calculate accessibility indicators, which we detail in the methodology section as they relate directly to measuring shared e-scooter effects on transport demand.

\section{Methodology}
\label{sec:methods}

To examine how e-scooters affect public transport use, we conducted a difference-in-differences analysis using geographical zones defined by DTPM as our spatial units. We divided urban areas into three groups using clustering to account for spatial differences and ensure valid comparisons between treatment zones (TAZs where shared e-scooter trips were recorded) and control zones (TAZs with no e-scooter activity). This section details our variable selection and methodology.

\subsection*{Selection of Study Variables}

We compiled an initial list of variables that are typically considered to influence travel behavior and assessed their relevance in the context of Santiago and this study. The candidate variables include sociodemographic variables from the 2017 census, such as \emph{underage population, working population, retired population}, \emph{percentage of female population}, \emph{average educational level}, and \emph{population density}. Additionally, we included variables from ADATRAP dataset to capture travel behavior, such as the \emph{average number of public transport trips per person}, \emph{average time of the first public transport trip} made each day, and \emph{average public transport trip distance}. We also included variables to determine the accessibility of each zone to public transport, counting bus stops per zone together with the derived bus-stop density and bus stops per 1,000 inhabitants, using data from OpenStreetMap and the General Transit Feed Specification (GTFS). These indicators were calculated for each zone in our design. A full description of these variables is provided in Table~B.10 in the Supplementary Material, along with spatial distribution maps (Figures~B.11 and~B.12)

\subsection*{Clustering and Definition of Treatment and Control Groups}

In this study, we aimed to create meaningful spatial clusters that reflect both statistical similarities and geographic proximity. The process of creating regions, known as regionalization ~\citep{duque2007supervised}, is crucial in this context. Regionalization is a specific form of clustering aimed at grouping observations that are alike in both their statistical characteristics and spatial proximity. This ensures that the resulting zones are not isolated but are similar to their neighboring areas, aligning with our goal of meaningful spatial clustering. 

We used Moran's I to analyze the spatial autocorrelation of the 12 candidate variables and retained the 10 exhibiting substantial spatial clustering (Moran's I above 0.2, all significant at $p \leq 0.003$ under 999 permutations); the two excluded variables (number of bus stops and bus stops per 1{,}000 people) showed weak spatial structure (Moran's I $\leq 0.13$). This ensured that the selected variables exhibited statistically significant and substantial spatial dependencies rather than random spatial distribution.

Based on these criteria, we selected the following cluster variables: \emph{percentage of underage population}, \emph{percentage of working population}, \emph{percentage of retired population}, \emph{average educational level}, \emph{percentage of female population}, \emph{bus stop density}, \emph{population density}, \emph{average number of public transport trips made per person}, \emph{average time of the first public transport trip of the day}, and \emph{average public transport trip distance}.

We implemented two clustering algorithms using Scikit-learn, a free, open-source Python library that provides tools for machine learning applications: k-means~\citep{lloyd1982least} and agglomerative hierarchical clustering. For agglomerative hierarchical clustering, we used the Ward method as the linkage criterion~\citep{ward1963hierarchical}, which minimizes the total within-cluster variance when merging clusters. We employed two different connectivity approaches for this method: a k-nearest neighbors (KNN) contiguity matrix and a sparse connectivity
matrix. Each approach was applied across a range of 2 to 10 clusters.

\begin{figure}
  \centering
  \includegraphics[width=0.95\linewidth]{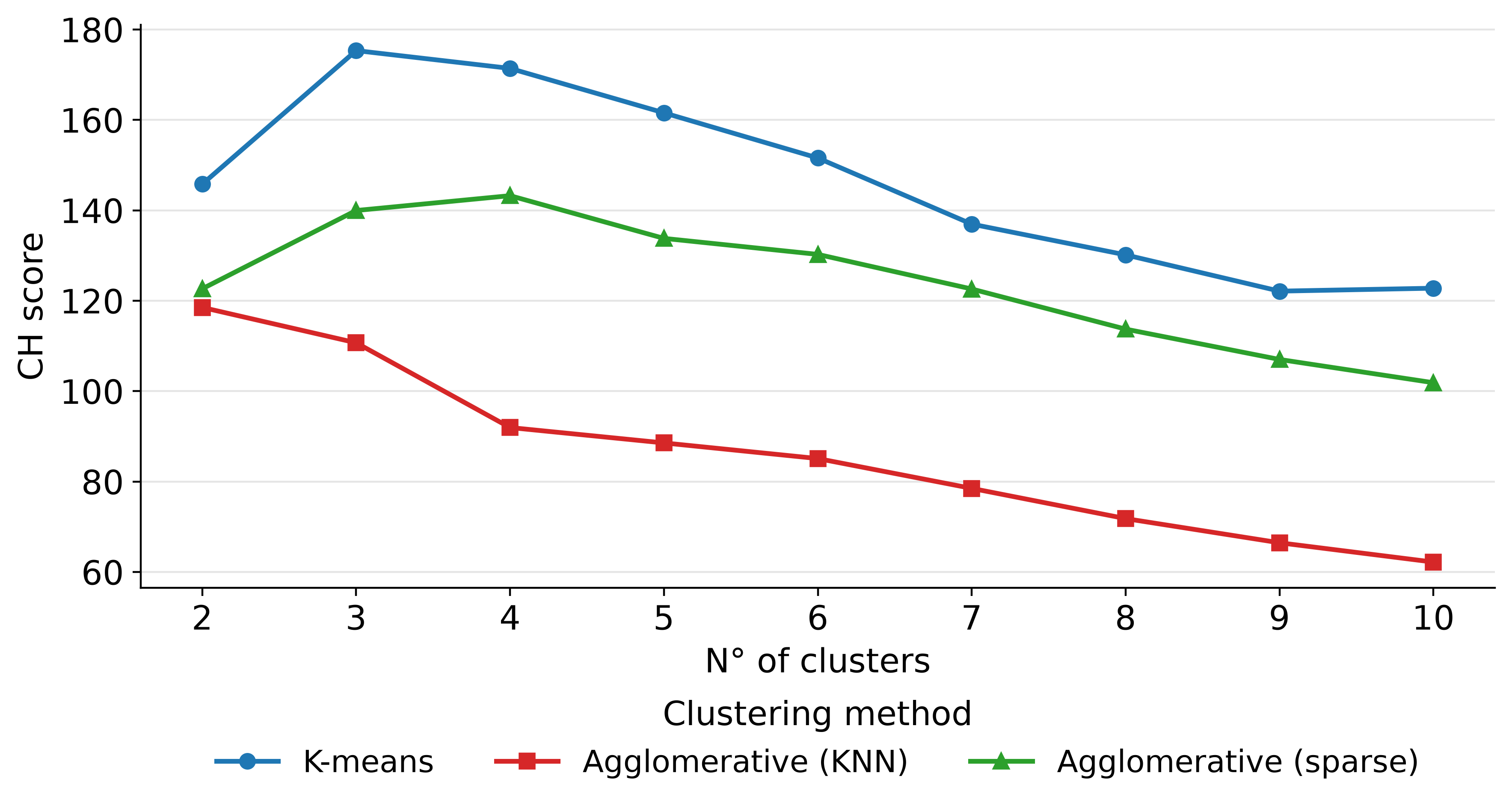}
  \caption{This figure shows the Calinski-Harabasz (CH) scores versus the
    number of clusters for three clustering methods: k-means
    (\emph{blue circles}), agglomerative hierarchical with KNN connectivity
    (\emph{red squares}), and agglomerative hierarchical with sparse connectivity
    \emph{(green triangles}). The x-axis represents the number of clusters (2 to 10),
    and the y-axis shows the CH score.}
\label{fig:ch}
\end{figure}

Afterward, we used the Calinski-Harabasz score~\citep{calinski1974dendrite} to compare the performance of these techniques and configurations to determine the best algorithm. The Calinski-Harabasz (CH) index, also known as the variance ratio criterion, measures the ratio of separation (the distance between cluster centroids) to cohesion (the distance between points within a cluster and their centroid). A higher score indicates
better-defined clusters. As illustrated in Figure \ref{fig:ch}, the k-means algorithm with 3 clusters achieved the highest score, making it our chosen model.

\begin{figure}
\centering
\includegraphics[width=0.95\linewidth]{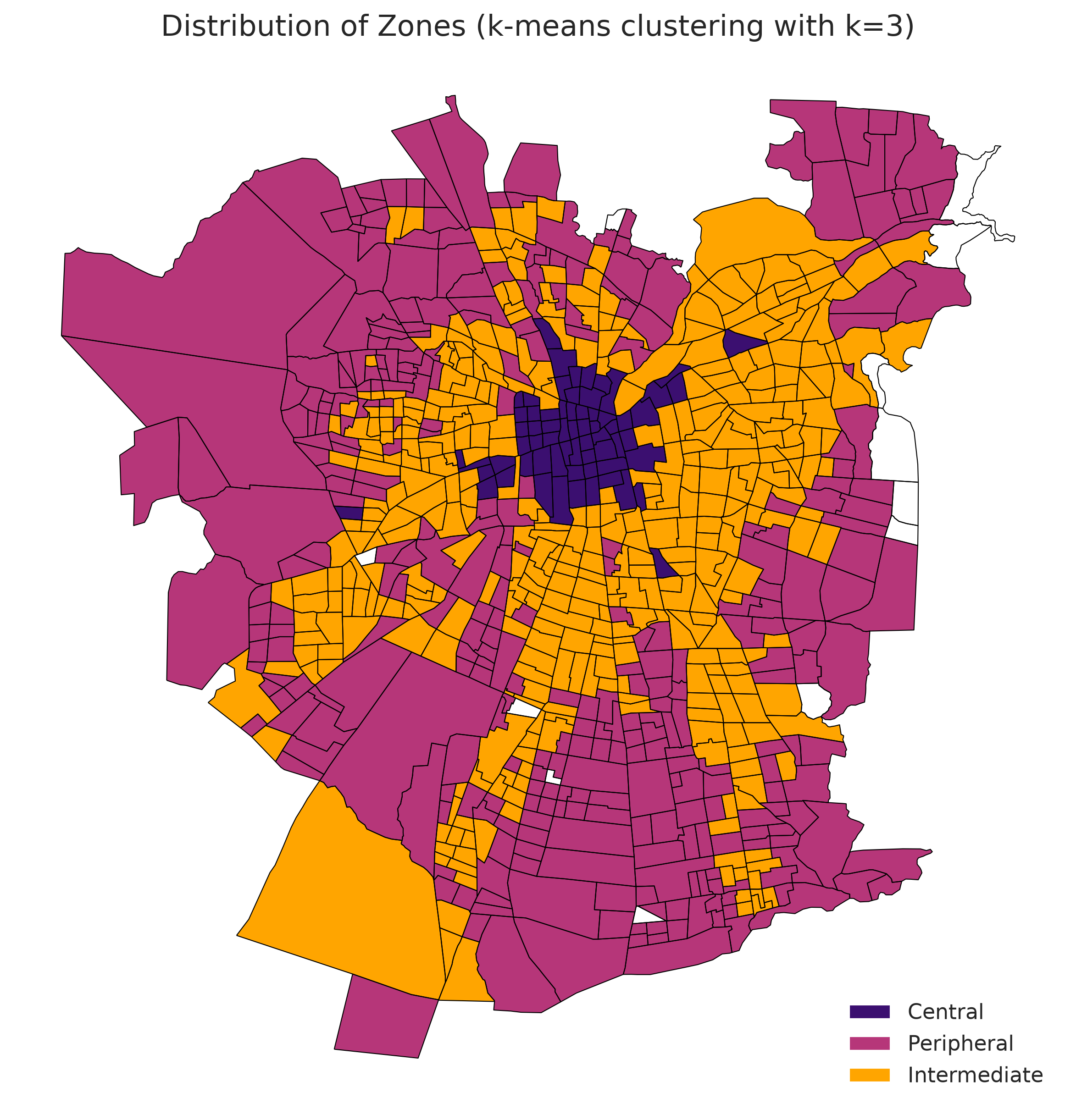}
\caption{Spatial distribution of zones resulting from the k-means
 clustering algorithm with k=3. The zones in \textit{dark purple} represent the
\emph{Central Region}, the zones in \textit{orange} correspond to an intermediate
area of the city (\emph{Intermediate Region}), and the zones in
\emph{magenta} represent the \emph{Peripheral Region}.}
\label{fig:clusters_reg}
\end{figure}

The resulting regionalization (groups of zones) after applying the
k-means clustering algorithm with k=3 is presented in Figure \ref{fig:clusters_reg}. The
three clusters, named according to their location in the city, almost
form concentric circles over Santiago. The zones in dark purple
represent the central area (\emph{Central Region}), the zones in magenta
represent the periphery (\emph{Peripheral Region}), and the zones in
orange correspond to an intermediate area of the city
(\emph{Intermediate Region}).

\begin{figure}[h]
\centering
\includegraphics[width=0.95\linewidth]{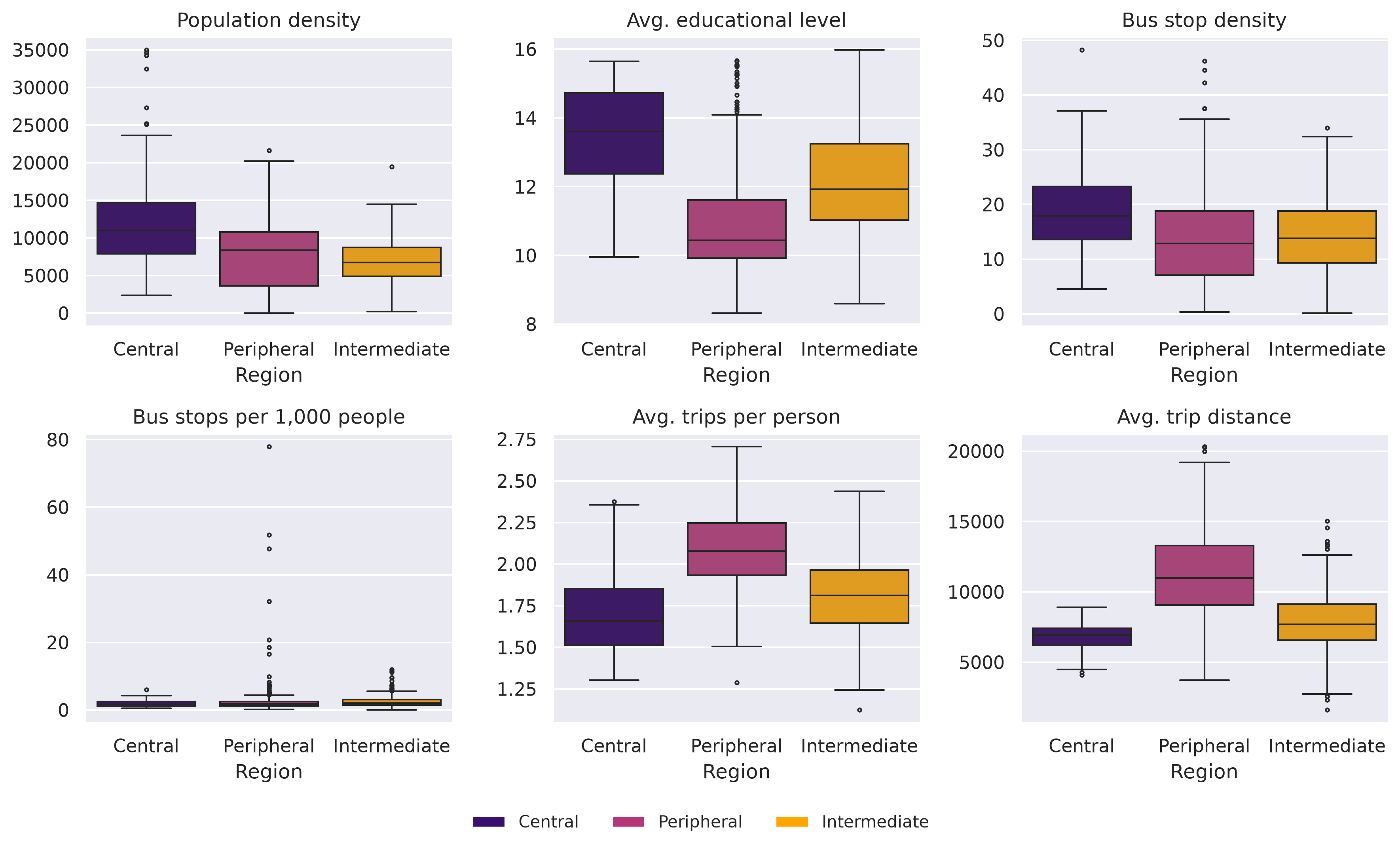}
\caption{Key demographic and transport characteristics across regions. Box plots of the six variables most relevant to the cluster interpretation: population density, average educational level, bus stop density, bus stops per 1,000 people, average number of trips per person, and average trip distance. Colours denote regions: \emph{Central} (dark purple), \emph{Peripheral} (magenta), and \emph{Intermediate} (orange). The complete set of candidate variables (12 demographic and transport metrics, of which 10 were retained for clustering) is reported in the Supplementary Material (Appendix~B).}
\label{fig:boxplot_key}
\end{figure}

After using the k-means algorithm to divide the city into three regions, we measured how the study variables were spread out in each region. This analysis revealed distinct characteristics for each region. Figure~\ref{fig:boxplot_key}
illustrates the characteristics of three distinct regions (\emph{Central, Peripheral}, and \emph{Intermediate}) in terms of demographic and transportation metrics. The \emph{Central Region} (\emph{dark purple}) has the highest population density, the highest average educational level and the highest bus-stop density, together with the largest share of working-age population and the smallest share of underage population. It is a dense, predominantly adult and highly educated core; its public-transport trips are the shortest of the three regions and the least frequent (fewer than two trips per person per day). The \emph{Peripheral Region} (\emph{magenta}) has the highest share of underage population and the lowest educational attainment, and its public-transport trips are the longest and most frequent; its population density is the second highest, after the Central Region. The \emph{Intermediate Region} (\emph{orange}) has the highest shares of retired and female population and takes intermediate values for population density, educational level and trip length, consistent with a transitional character between the Central and Peripheral regions. In terms of transport usage, the Peripheral Region shows the highest average number of trips per person and the longest trips, whereas the Central Region shows the shortest trips; bus-stop density, and hence accessibility to public transport, is highest in the Central zone.

\begin{figure}[H]
\centering
\includegraphics[width=\linewidth]{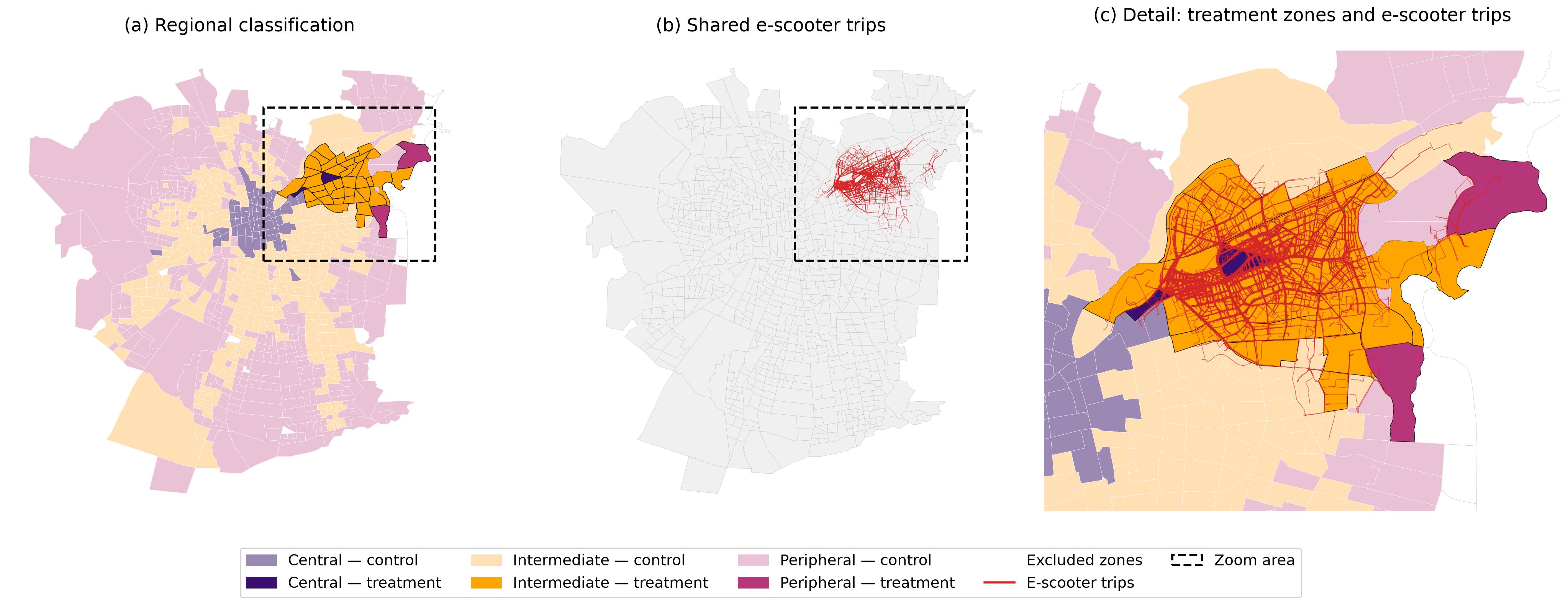}
\caption{Spatial distribution of e-scooter trips and regional classification of treatment zones. Panel~(a) shows the regional classification of all traffic analysis zones across Santiago, with darker shades indicating treatment zones and the dashed rectangle marking the e-scooter service area. Panel~(b) shows the geographic distribution of shared e-scooter trips across the city. Panel~(c) provides a zoomed view of the service area, overlaying trip trajectories on the regional classification to illustrate which regions contain treatment zones. The e-scooter service area falls primarily within the Intermediate region, with a small number of Central treatment zones at its western edge.}
\label{fig:scooters}
\end{figure}

After applying the clustering algorithm and dividing the zones into three groups of regions, we excluded 56 zones from the study. Although these zones had no origin and destination trips involving scooters, they were close to the treatment zones and therefore could still experience indirect mobility impacts post-intervention, potentially contaminating our control group. We excluded zones intersecting a 1,440 m buffer applied to each scooter trajectory, a distance chosen based on the average trip distance calculated from our data. Figure~\ref{fig:scooters} displays the spatial distribution of e-scooter trips, treatment zones, and their regional classification across the study area, along with Table \ref{table:regions}. Because treatment is defined by the direction of e-scooter flows, a zone may qualify as treated for the generation model (e-scooter trips arriving in the zone) but not for the attraction model (trips departing from it); Table~\ref{table:regions} therefore reports both counts.

\begin{table}[H]
\centering
\scriptsize
\begin{tabularx}{\textwidth}{@{}l *{5}{>{\centering\arraybackslash}X}@{}}
\toprule
\textbf{Region} & \textbf{Total zones} & \textbf{Excluded (buffer)} & \textbf{Treatment (generation)} & \textbf{Treatment (attraction)} & \textbf{Control} \\\midrule
Central Region & 66 & 8 & 2 & 2 & 56 \\
Intermediate Region & 388 & 27 & 38 & 38 & 322 \\
Peripheral Region & 337 & 17 & 2 & 2 & 318 \\\bottomrule
\end{tabularx}
\caption{Number of zones by region and study role. A zone is treated for the trip generation model if e-scooter trips end in it (at least 5 trips per day) and for the trip attraction model if e-scooter trips start in it, so the two sets of treated zones can differ where e-scooter flows are asymmetric. In the Intermediate Region, 37 zones are treated in both models and one more in each, so each model compares 38 treated zones with the other 323 zones outside the buffer. \emph{Control} zones are those not treated in either model and located outside the 1,440~m buffer around scooter trajectories (\emph{Excluded}); the four remaining buffer zones fall outside the three regions. Totals refer to the 791 zones with complete data used in the clustering.}
\label{table:regions}
\end{table}

\subsection*{Regression Model}

Finally, to measure how the implementation of e-scooters affects public transportation demand, we applied a difference-in-differences (DiD) approach in conjunction with two distinct Negative Binomial (NB) regression models \citep{cameron1986econometric}. The first model evaluates trip generation, focusing on trips originating from public transport stations, while the second evaluates trip attraction, analyzing trips ending at public transport stations. These models were chosen because public transportation trips often exhibit significant variability and over-dispersion, where the variance exceeds the mean, making traditional Poisson regression unsuitable for our dataset. The NB models effectively address over-dispersed count data, allowing us to better capture the variability in public transportation trips across different zones and providing a more accurate understanding of the factors influencing both trip generation and attraction scenarios. Because treatment is assigned to zones and does not vary within a zone and period, the models are fitted to the total trips of each zone in each period, so that the unit of analysis coincides with the unit of treatment \citep{bertrand2004much}.

The choice of NB models aligns with previous studies that have employed similar approaches to analyze transportation demand across various modes \citep[e.g.,][]{schimohr2021spatial,hu2022understanding,manville2023vehicle}. It is also consistent with prior research examining behavioral and spatial patterns in Santiago \citep{graells2017effect,beiro2018shopping,graells2023measuring}.

The DiD methodology estimates effects by comparing changes in outcomes over time between treatment and control groups before and after an intervention. The treatment group comprises the TAZs where shared e-scooter trips were observed; the control group includes zones without e-scooter trips and located outside a 1,440-meter buffer around scooter trajectories, which minimizes indirect influences from nearby e-scooter activity on control zones. Applying this distinction within the regional clusters defined earlier ensures that comparisons between treatment and control zones reflect differences associated with e-scooter introduction, rather than variations in sociodemographic or transport characteristics.

For the difference-in-differences design, we established a minimum threshold of 5 daily e-scooter trips per zone as the treatment assignment criterion. This threshold ensures that only zones with sufficient e-scooter activity are considered "treated" by the intervention. Based on empirical analysis of zone-day observations for both e-scooter origins (300 observations) and destinations (323 observations), the 5-trip threshold is positioned above the 25th percentile (2.0 trips/day for both) and at or below the median (5.0 trips/day for origins, 6.0 trips/day for destinations). This results in 53.3\% of origin observations and 54.2\% of destination observations being assigned to treatment, with consistent assignment rates across flow types (0.9 percentage point difference). The threshold effectively excludes zones with sporadic activity that would not be meaningfully exposed to e-scooter interventions, preserving the internal validity of the difference-in-differences design. A sensitivity analysis re-estimating the main models under alternative thresholds (1, 3, 5, 8, 10, and 15 trips per day) is reported in the Supplementary Material (Appendix~I).

For the trip generation model, we examine how e-scooters arriving at a zone (destinations) affect the generation of public transport trips from that same zone, evaluating whether e-scooters function as feeders that bring people to transit stations. Conversely, for the trip attraction model, we analyze how e-scooters departing from a zone (origins) influence public transport trips ending in that zone, assessing whether e-scooters serve as last-mile connections from transit stations.

\begin{equation}
\begin{aligned}
\ln\left({\textbf{E}}[y_{gen} \mid \mathbf{x}_{gen}] \right) = & \, \beta_0 + \beta_1 \text{treatment\_zones} + \beta_2 \text{post} \\
& + \beta_3 (\text{treatment\_zones} \times \text{post}) \\
& + \boldsymbol{\beta} \cdot \text{Controls} + \epsilon,
\end{aligned}
\end{equation}

\begin{equation}
\begin{aligned}
\ln\left({\textbf{E}}[y_{att} \mid \mathbf{x}_{att}] \right) = & \, \beta_0 + \beta_1 \text{treatment\_zones} + \beta_2 \text{post} \\
& + \beta_3 (\text{treatment\_zones} \times \text{post}) \\
& + \boldsymbol{\beta} \cdot \text{Controls} + \epsilon,
\end{aligned}
\end{equation}

where $y_{gen}$ and $y_{att}$ represent the count of public transport trips in a TAZ for trip generation and attraction scenarios, respectively. The variable
\textit{treatment\_zones} is a dummy variable indicating whether the zone is associated with e-scooter trips: for the trip generation model, it takes the value 1 if e-scooter trips end in the zone (destinations), and for the trip attraction model, it takes the value 1 if e-scooter trips start in the zone (origins). The variable \textit{post} is a dummy indicating the post-implementation period (2019), and the interaction term \textit{treatment×post} captures the difference-in-differences (DiD) effect.

The \(\boldsymbol{\beta} \cdot \text{Controls}\) term incorporates additional control variables to account for variations in trip counts due to socio-demographic and spatial factors. These control variables include the log-transformed count of the working population in the TAZ, the average educational level relative to the regional average, and an indicator of whether the zone had a metro station within a 750-meter radius before the opening of Metro Line 3. In addition, we control for confounding metro infrastructure changes between the pre- and post-treatment periods: an indicator for the zones within 750~m of a Metro Line 3 station (opened January 22, 2019) and its interaction with the post period (\texttt{L3\_post}), and analogous terms for the zones that gained their first metro station within 750~m; an analogous control for Metro Line 6 (opened November 2017), \texttt{L6\_post}, is used in the placebo comparison between 2017 and 2018. These controls separate the estimates from concurrent metro expansion; a robustness analysis of these specifications is provided in the Supplementary Material (Appendix~G).

Because the negative binomial model uses a logarithmic link, its coefficients act multiplicatively on the expected trip count: for any covariate, $\exp(\beta)-1$ is the proportional change in expected trips associated with that covariate. The difference-in-differences estimate is therefore $\exp(\beta_3)-1$, the percentage change in expected public-transport trips in treated zones, relative to control zones, associated with the introduction of e-scooters; all impact magnitudes reported in Section~\ref{sec:results} are computed this way and are thus independent of baseline trip levels. A positive $\beta_3$ is consistent with complementarity (e-scooters are associated with more public-transport trips), and a negative $\beta_3$ with substitution. The model also estimates a dispersion parameter that accommodates the over-dispersion noted above; in every specification this parameter was significantly greater than zero, confirming the negative binomial model over a Poisson alternative.

Standard errors are clustered by zone. Because cluster-robust inference is unreliable when few clusters are treated, as in the Central and Peripheral regions, we also report randomization-inference $p$-values \citep{conley2011inference,mackinnon2020randomization}: treatment is reassigned at random among the zones of the region, keeping the number of treated zones fixed, and the observed $t$-statistic is compared with its distribution over reassignments (5,000 random reassignments, or all possible assignments when there are fewer). Confidence intervals are based on the clustered standard errors and are reported for the cells with at least five treated zones; statistical significance is assessed with the randomization-inference $p$-values (see the Supplementary Material, Appendix~H).

For each scenario we estimated three separate models, classifying every trip by the mode of its \emph{boundary} stage: the boarding (first) stage for trip generation and the alighting (last) stage for trip attraction. The three models are: (1) \textbf{Bus}, trips whose boundary stage is a bus service (including pre-paid \emph{zona paga} boardings); (2) \textbf{Metro}, trips whose boundary stage is metro or commuter rail (Metrotr\'en); and (3) \textbf{Bus+Metro}, the union of the two, that is, all public-transport trips whose boundary stage is bus or metro. In every model the dependent variable is the count of trips in each Traffic Analysis Zone (TAZ), and each trip contributes a \emph{single} count to the zone where it boards (generation) or alights (attraction); multi-stage journeys are therefore not counted per leg nor double-counted across models. The Bus+Metro scenario is thus an \emph{aggregate} of public-transport demand (bus or metro), not a measure of multimodal journeys. We applied each model to the three regions (\textit{Central}, \textit{Intermediate}, and \textit{Peripheral}), each containing its respective control and treatment areas.

To capture multimodal journeys explicitly, we additionally estimated a \emph{strict-intermodal} specification that isolates trips genuinely combining bus and metro within a single journey: trips whose first stage is bus and any other stage is metro or commuter rail (Metrotr\'en), or whose first stage is metro or commuter rail and any other stage is bus. By definition these journeys involve at least two stages, so this model uses a dependent variable distinct from the combined Bus+Metro model and tests whether bus--metro combinations, rather than single-mode trips, change in zones with e-scooter activity. The smart-card and e-scooter records cannot be linked to the same traveler or journey, so first/last-mile connectivity is inferred rather than observed. We fit it for both trip generation and trip attraction within the same difference-in-differences framework (Tables~\ref{tab:intermodal_generation}--\ref{tab:intermodal_attraction}).

\begin{figure}[h]
	\centering
	\includegraphics[width=0.95\linewidth]{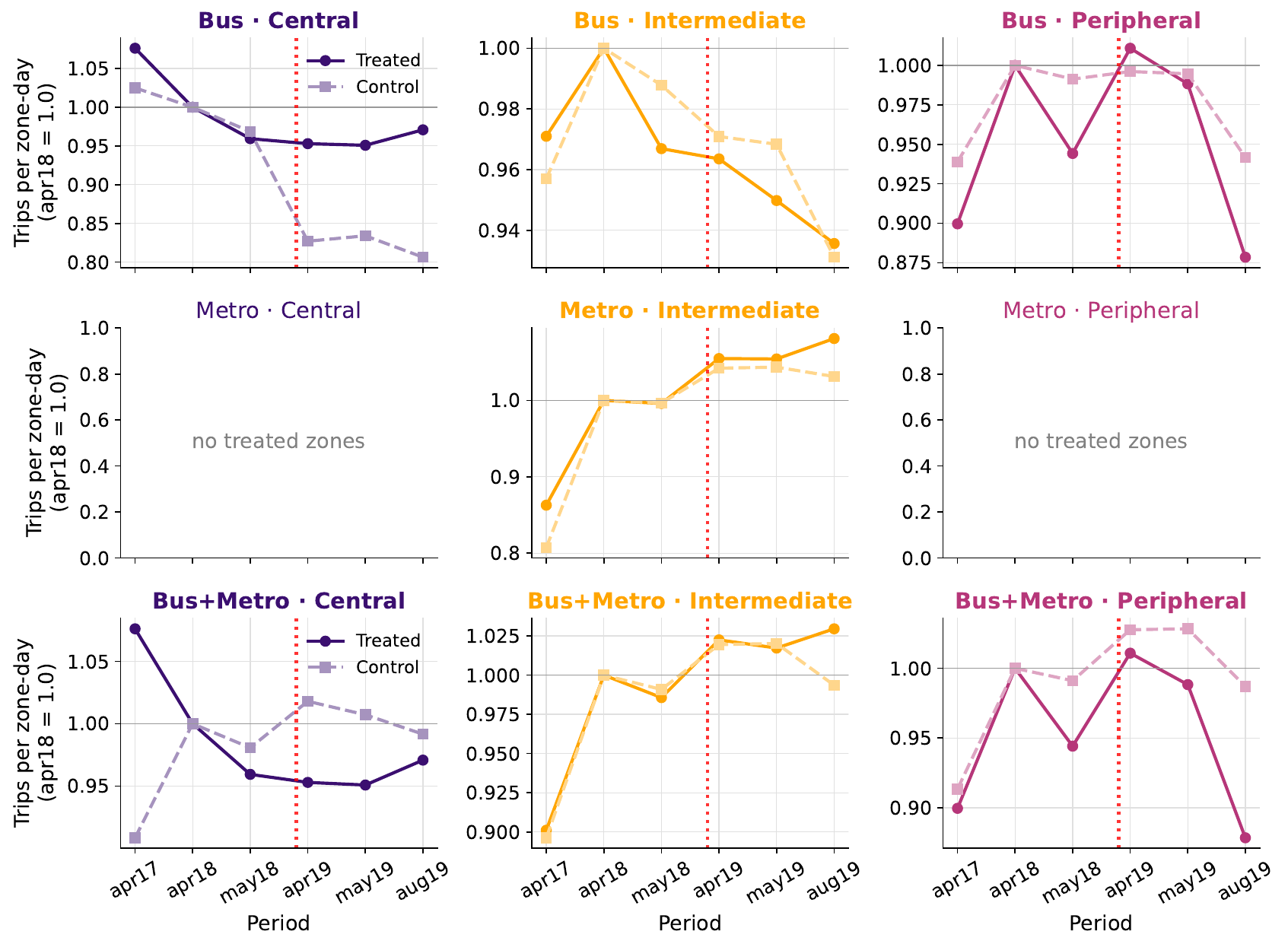}
	\caption{Trajectories of public-transport trip generation in treated and control zones. Public-transport trips per zone-day, normalised to April 2018 = 1.0, for treated and control zones across nine model $\times$ region panels. In the Intermediate Region, treated and control series move closely together over the pre-treatment span (April 2017 -- May 2018) for the Bus+Metro model, with a small gap between April and May 2018; for the Metro model they co-move between April and May 2018 but differ in April 2017, and for the Bus model they separate between April and May 2018. The Central and Peripheral panels are based on two treated zones each, and the Metro panels of these regions are empty because no treated zone has metro boardings. The vertical line marks March 2019, when shared e-scooter usage reached operational maturity.}
	\label{fig:parallel_trends_methods}
\end{figure}

Identification relies on the parallel-trends assumption: in the absence of the e-scooter introduction, treated and control zones would have exhibited similar trends in public-transport demand. Because the main design relies on only two time points, we examine this assumption empirically with trip counts for additional windows (April 2017, May 2018, May 2019 and August 2019) and refit the model to a series of placebo specifications that compare periods with the same e-scooter status. For the Metro and intermodal models in the Intermediate Region, the placebo estimates provide supporting evidence for the assumption; the Bus model is the exception, as the pre-treatment placebo indicates diverging trends, which also appear, attenuated, in the combined Bus+Metro model, so Bus-specific estimates should be interpreted with caution. These tests do not establish the assumption, particularly where few zones are treated, and the available windows do not allow a multi-period event-study assessment. In addition, because treatment is defined by observed e-scooter use rather than by service availability, the estimates describe changes in the zones where the service was used and may partly reflect the selection of those zones; an alternative definition based on service availability, in which all zones of the municipality where the service started are treated regardless of use, leads to the same conclusions (Supplementary Material, Appendix~L). Figure~\ref{fig:parallel_trends_methods} shows the normalised trajectories across the six windows; full numerical results and the corresponding tests are reported in the Supplementary Material (Appendix~F).

\section{Results}
\label{sec:results}

In this section, we present our findings on the impact of shared e-scooter introduction on public transportation usage in Santiago, Chile. Using negative binomial regression with a difference-in-differences approach, we analyze how the deployment of e-scooter services affects public transport demand across urban regions. Our analysis distinguishes between trip generation (public transport boardings - trips originating in each zone) and attraction (public transport alightings - trips ending in each zone) to capture the multidirectional effects of this new micro-mobility service.

\subsection*{Generation of Trips}\label{first-mile-trips}
\addcontentsline{toc}{subsection}{Generation trips}

The full model estimates for public transport trip generation are reported in Appendix D. Table \ref{tab:escooter_impact} summarizes the marginal effects of shared e-scooter introduction on trip generation across regions, and Table \ref{tab:trips_average} reports the average public transport trips by region and mode, providing baseline context for interpreting the percentage changes.

We examine how e-scooter trips arriving at each zone influence the generation of public transport trips from those same zones. This analysis examines whether e-scooters function as feeders to public transport (bringing people to transit stations who then continue their journeys via bus or metro) or whether they operate independently. Our findings show no statistically significant change in total trip generation in the region that concentrates the service.

In the Central Region, where only two zones are treated, total public transport boardings decrease by $5.7\%$ ($p=0.023$ under randomization inference), while bus boardings show no significant change ($+1.0\%$, $p=0.88$) and no treated zone has metro boardings. Because this estimate rests on two zones, whose boardings had already changed by $-19.0\%$ relative to the control zones between April 2017 and April 2018, we do not interpret it as an effect of e-scooters (Appendix~H).

In the Intermediate Region, which contains 38 of the 42 treated zones, total public transport boardings changed by $-0.9\%$ (95\% CI: $-3.3$, $+1.6$; $p=0.60$). The estimate is not statistically significant, and its confidence interval excludes changes larger than about 4\% in either direction. Metro boardings show no significant change either ($+0.5\%$; $-2.6$, $+3.7$; $p=0.87$). Bus boardings show a decrease ($-2.9\%$; $-4.9$, $-0.8$; $p=0.009$), which we do not interpret as an effect of e-scooters because a difference of similar size appears between April and May 2018, before their introduction ($-2.5\%$, $p<0.001$; Appendix~F).

In the Peripheral Region, also with two treated zones, neither the combined bus and metro effect ($+1.1\%$, $p=0.69$) nor the bus effect ($+1.7\%$, $p=0.53$) is statistically significant. Metro boardings could not be evaluated for this region because no treated zone has metro boardings.

The absence of a statistically significant effect on total boardings is consistent with the scale of e-scooter use. On an average working day of the study week, 623 e-scooter trips ended in the treated zones of the Intermediate Region, against 400,664 public transport boardings in the same zones (0.16\%). Even if every e-scooter trip had replaced, or generated, one public transport trip, total boardings would have changed by less than 0.2\%.

Whether e-scooter activity is associated with changes in genuine bus--metro combinations, rather than in single-mode trips, is examined with a strict-intermodal model in the Intermodal Trips subsection below.

\begin{table*}[t]
	\centering
	\scriptsize

\begin{tabularx}{\textwidth}{@{}p{2.4cm} X X X@{}}
	\toprule
	\textbf{Mode of Transport} & \textbf{Central} & \textbf{Intermediate} & \textbf{Peripheral} \\
	\midrule
	\textbf{Bus and Metro}
	& \textbf{$-$5.7\%**} ($p=0.023$) & $-$0.9\% [$-3.3$, $+1.6$] ($p=0.596$) & +1.1\% ($p=0.694$) \\
	\midrule
	\textbf{Bus}
	& +1.0\% ($p=0.882$) & \textbf{$-$2.9\%***} [$-4.9$, $-0.8$] ($p=0.009$) & +1.7\% ($p=0.526$) \\
	\midrule
	\textbf{Metro}
	& -- & +0.5\% [$-2.6$, $+3.7$] ($p=0.865$) & -- \\
	\bottomrule
\end{tabularx}
	\caption{Difference-in-differences estimates of e-scooter impact on public transport trip generation (boardings). Percentage changes (Treatment $\times$ Post) from negative binomial regression models fitted to zone totals, with controls for spatial and demographic factors and time-variant indicators for the opening of Metro Line 3. In brackets, 95\% confidence intervals from standard errors clustered by zone, reported for cells with at least five treated zones; in parentheses, randomization-inference $p$-values. Asterisks denote significance under randomization inference: * $p<0.10$; ** $p<0.05$; *** $p<0.01$. Metro cannot be estimated in the Central and Peripheral regions. Baseline: April 9--13, 2018 $\leftrightarrow$ April 8--12, 2019 (Monday--Friday).}
\label{tab:escooter_impact}
\end{table*}

\begin{table*}[t]
	\centering
	\scriptsize
	\begin{tabular}{@{}lrrrrrr@{}}
		\toprule
		\multirow{2}{*}{\textbf{Region}} & \multicolumn{3}{c}{\textbf{Boardings (trips/hr/zone)}} & \multicolumn{3}{c}{\textbf{Alightings (trips/hr/zone)}} \\
		\cmidrule(lr){2-4} \cmidrule(lr){5-7}
		& \textbf{Bus} & \textbf{Metro} & \textbf{Bus or Metro} & \textbf{Bus} & \textbf{Metro} & \textbf{Bus or Metro} \\
		\midrule
		Central      & 239.40 & 1546.99 & 696.16 & 210.63 & 1721.61 & 791.45 \\
		Intermediate & 101.63 & 593.02 & 169.44 & 93.86 & 642.48 & 198.46 \\
		Peripheral   & 94.38 & -- & 108.22 & 75.47 & -- & 105.24 \\
		\bottomrule
	\end{tabular}
	\caption{Baseline public transport demand by region and mode. Values represent average trips per hour per zone during the pre-treatment period (April 2018), providing context for interpreting percentage changes in Tables \ref{tab:escooter_impact} and \ref{tab:attraction}. Boardings capture trip origins (generation model) while alightings capture trip destinations (attraction model). Baseline averages are computed on the April 9--13, 2018 (Monday--Friday) data. Metro figures for the Peripheral Region are omitted, consistent with the model tables, as that cluster lacks sufficient metro coverage for estimation.}
	\label{tab:trips_average}
\end{table*}

\subsection*{Attraction of Trips}
The full difference-in-differences estimates for public transport trip attraction are reported in Appendix E. Table \ref{tab:attraction} summarizes the marginal effects of e-scooter introduction on public transport trips ending in each region, reporting the percentage change and significance for every transport mode and region.

\begin{table*}[t]
	\centering
	\scriptsize
	\begin{tabularx}{\textwidth}{@{}p{2.4cm} X X X@{}}
		\toprule
		\textbf{Mode of Transport} & \textbf{Central} & \textbf{Intermediate} & \textbf{Peripheral} \\
		\midrule
		\textbf{Bus and Metro}
		& \textbf{$-$3.6\%*} ($p=0.082$) & $-$1.1\% [$-3.8$, $+1.6$] ($p=0.523$) & +12.1\% ($p=0.233$) \\
		\midrule
		\textbf{Bus}
		& +1.7\% ($p=0.711$) & \textbf{$-$3.5\%**} [$-6.5$, $-0.4$] ($p=0.032$) & +12.6\% ($p=0.216$) \\
		\midrule
		\textbf{Metro}
		& -- & $-$10.8\% [$-36.0$, $+24.2$] ($p=0.559$) & -- \\
		\bottomrule
	\end{tabularx}
	\caption{Difference-in-differences estimates of e-scooter impact on public transport trip attraction (alightings). Percentage changes (Treatment $\times$ Post) from negative binomial regression models fitted to zone totals, with controls for spatial and demographic factors and time-variant indicators for the opening of Metro Line 3. In brackets, 95\% confidence intervals from standard errors clustered by zone, reported for cells with at least five treated zones; in parentheses, randomization-inference $p$-values. Significance under randomization inference: * $p<0.10$; ** $p<0.05$; *** $p<0.01$. Metro cannot be estimated in the Central and Peripheral regions.}
	\label{tab:attraction}
\end{table*}

The attraction model shows the same pattern as trip generation. In the Intermediate Region, the combined Bus+Metro effect is $-1.1\%$ (95\% CI: $-3.8$, $+1.6$; $p=0.52$), and the estimate for Metro alightings is imprecise and not statistically significant ($-10.8\%$; $-36.0$, $+24.2$; $p=0.56$). On an average working day, 615 e-scooter trips started in the treated zones of this region, against 425,149 public transport alightings (0.14\%).

Bus alightings in the Intermediate Region decrease ($-3.5\%$; $-6.5$, $-0.4$; $p=0.032$). As for boardings, however, the pre-treatment placebo is also negative ($-2.5\%$, $p=0.002$), and the estimate obtained with data for May is smaller and not significant ($-1.3\%$, $p=0.50$; Appendix~F), so we do not interpret this decrease as an effect of e-scooters.

In the Central and Peripheral regions, the estimates for combined Bus+Metro alightings are $-3.6\%$ ($p=0.08$) and $+12.1\%$ ($p=0.23$), respectively. With two treated zones in each region these estimates are not informative, and we do not interpret them.

\subsection*{Intermodal Trips}
\addcontentsline{toc}{subsection}{Intermodal Trips}

Tables \ref{tab:intermodal_generation} and \ref{tab:intermodal_attraction} present the difference-in-differences estimates for trips that strictly combine bus and metro within a single journey. Unlike the combined Bus+Metro model in Tables \ref{tab:escooter_impact} and \ref{tab:attraction}, which captures any trip whose first or last stage uses bus or metro, the intermodal model isolates trips whose first stage is bus and any other stage is metro/metrotren, or whose first stage is metro/metrotren and any other stage is bus. This specification tests whether bus--metro combinations change in zones with e-scooter activity.

\begin{table}[h]
	\centering
	\scriptsize
	\begin{tabular}{p{5.2cm}ccc}
		\toprule
		\textbf{Model} & \textbf{Central} & \textbf{Intermediate} & \textbf{Peripheral} \\
		\midrule
		\textbf{Intermodal Bus+Metro (generation)} & +4.3\% & \textbf{$-$4.7\%***} & $-$1.2\% \\
		\quad 95\% confidence interval & -- & [$-7.4$, $-1.8$] & -- \\
		\quad $p$-value (randomization inference) & 0.780 & 0.004 & 0.621 \\
		\quad Treated zones / zones in the model & 2/58 & 38/359 & 2/320 \\
		\quad Baseline: trips per day, treated zones & 1,128 & 79,613 & 5,708 \\
		\bottomrule
	\end{tabular}
	\caption{Difference-in-differences estimates for intermodal trip \emph{generation}, with 95\% confidence intervals (zone-clustered standard errors; reported for cells with at least five treated zones), randomization-inference $p$-values, number of zones and pre-treatment baseline. Asterisks denote significance under randomization inference: * $p<0.10$; ** $p<0.05$; *** $p<0.01$. Apr 2018 $\leftrightarrow$ Apr 2019 baseline with L3 control.}
	\label{tab:intermodal_generation}
\end{table}

\begin{table}[h]
	\centering
	\scriptsize
	\begin{tabular}{p{5.2cm}ccc}
		\toprule
		\textbf{Model} & \textbf{Central} & \textbf{Intermediate} & \textbf{Peripheral} \\
		\midrule
		\textbf{Intermodal Bus+Metro (attraction)} & +1.6\% & \textbf{$-$5.7\%**} & +6.2\% \\
		\quad 95\% confidence interval & -- & [$-9.5$, $-1.8$] & -- \\
		\quad $p$-value (randomization inference) & 0.811 & 0.010 & 0.478 \\
		\quad Treated zones / zones in the model & 2/58 & 38/360 & 2/318 \\
		\quad Baseline: trips per day, treated zones & 588 & 106,196 & 2,218 \\
		\bottomrule
	\end{tabular}
	\caption{Difference-in-differences estimates for intermodal trip \emph{attraction}. Same specification as Table \ref{tab:intermodal_generation}.}
	\label{tab:intermodal_attraction}
\end{table}

\clearpage
\begin{figure}[H]
	\centering
	\includegraphics[width=\linewidth,height=0.8\textheight,keepaspectratio]{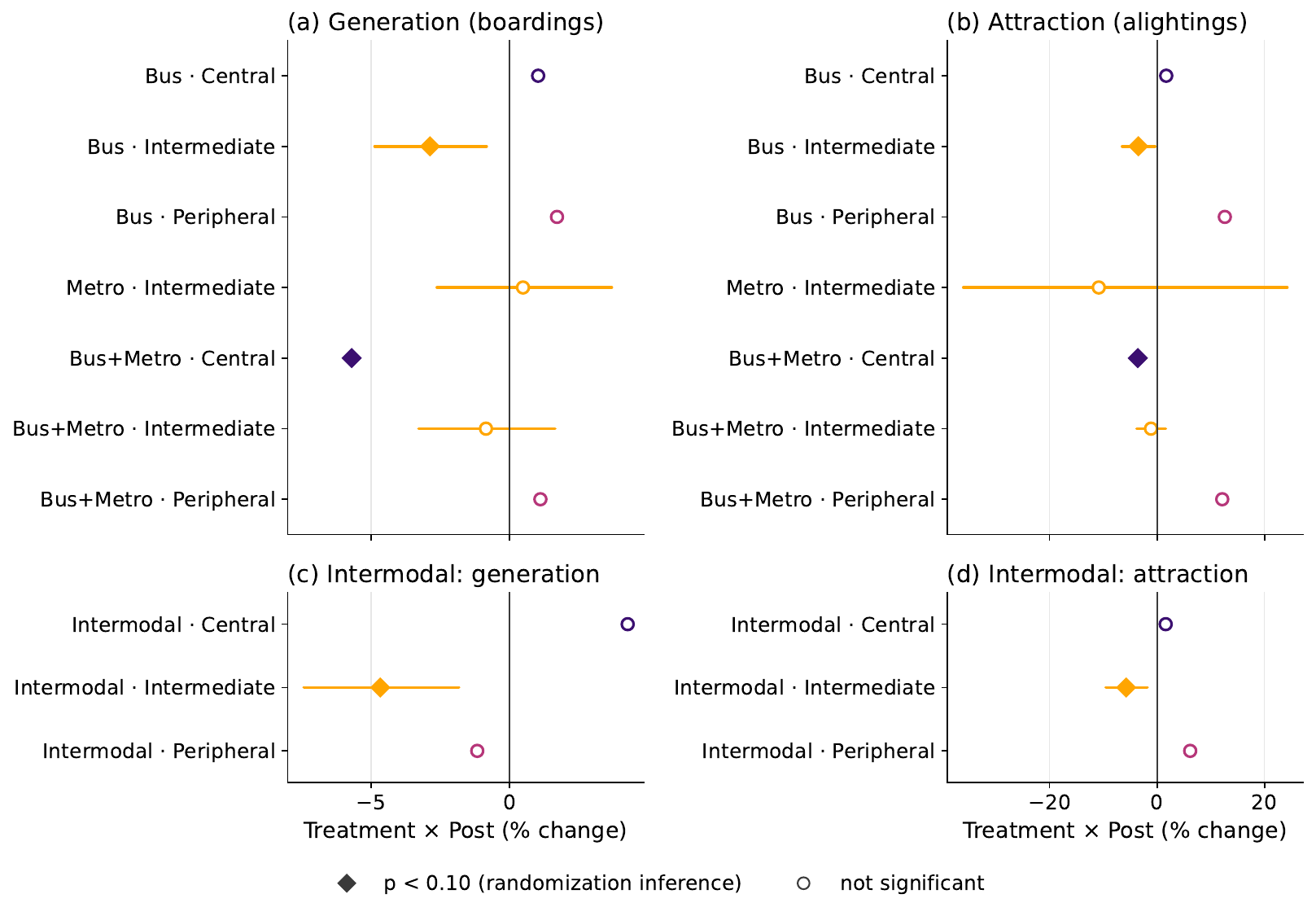}
	\caption{DiD effects of e-scooter introduction on public transport demand. Each panel shows Treatment $\times$ Post coefficients (percentage change); 95\% confidence intervals are shown for cells with at least five treated zones. Filled diamonds indicate significance at $p<0.10$ under randomization inference; hollow circles indicate non-significant estimates. Panels~(a) and~(b) show generation (boardings) and attraction (alightings) for all estimable models and regions (Metro can be estimated only in the Intermediate Region); panels~(c) and~(d) show the strict-intermodal specification (trips combining bus and metro within a single journey).}
	\label{fig:forest_all}
\end{figure}

Two findings stand out. First, intermodal generation in the Intermediate Region is negative and highly significant ($-4.7\%$; 95\% CI: $-7.4$, $-1.8$; $p=0.004$): trips that combine bus and metro within a single journey decrease in zones where e-scooters operate, relative to control zones. Second, intermodal attraction in the same region also decreases ($-5.7\%$; $-9.5$, $-1.8$; $p=0.010$). Since total boardings and alightings show no statistically significant change, these estimates indicate a change in the composition of trips rather than in their number. The intermodal estimates for the Central and Peripheral regions are not statistically significant.

This decline is consistent with a first/last-mile substitution mechanism, in which the short bus stage of a combined journey is replaced, but two considerations limit this interpretation. First, the smart-card and e-scooter records cannot be linked to the same traveler or journey, so the use of an e-scooter in place of a bus stage is inferred rather than observed. Second, the decline is larger than the observed e-scooter trips could account for: these trips are equivalent to 0.78\% of the bus--metro trips that start in the treated zones, and to 0.58\% of those that end there, below the lower limit of the confidence intervals (1.8\% in both cases). The decline is therefore associated with e-scooter activity, but it cannot be attributed to the observed e-scooter trips alone.

The association is stronger where e-scooter use is more intense. Dividing the treated zones of the Intermediate Region into terciles of daily e-scooter trips, the intermodal generation estimates are $-1.3\%$, $-3.1\%$ and $-7.8\%$ from the lowest to the highest tercile; with a continuous measure of intensity, the estimates are $-1.7\%$ per unit of the logarithm of daily e-scooter trips for generation ($p=0.001$) and $-2.1\%$ for attraction ($p=0.001$). The combined Bus+Metro model shows no such gradient (Appendix~J).

\subsection*{Robustness and Model Validation}
\addcontentsline{toc}{subsection}{Robustness and Model Validation}

We examine the robustness of our difference-in-differences estimates for the Intermediate Region, the only region with enough treated zones for this purpose, by re-estimating the combined Bus+Metro and intermodal models under alternative periods, definitions of treated zones, and model choices. Tables \ref{tab:robustness_generation} and \ref{tab:robustness_attraction} present the results for trip generation and attraction, respectively; complete results are reported in the Supplementary Material (Appendices~F, G and~I).

First, the results do not depend on the weeks compared. With data for May 2018 and May 2019, the combined Bus+Metro effect is $+0.0\%$ for generation and $-0.3\%$ for attraction, and the intermodal effect is $-4.7\%$ ($p=0.015$) and $-4.4\%$ ($p=0.075$). The placebo tests compare periods with the same e-scooter status. For intermodal trips they are close to zero: $-0.0\%$ and $-0.6\%$ between April and May 2018, before the introduction of e-scooters, and $-0.4\%$ and $+0.8\%$ between April and May 2019. For the combined Bus+Metro model, the placebo between April and May 2018 is small but negative ($-1.4\%$, $p=0.039$, for generation; $-1.6\%$, $p=0.078$, for attraction), which reflects the pre-existing trend in bus trips noted above; it is $-0.8\%$ between April 2017 and April 2018 and $-0.6\%$ in both directions between April and May 2019.

\begin{table}[h]
	\centering
	\scriptsize
	\begin{tabular}{@{}lcc@{}}
		\toprule
		\textbf{Specification} & \textbf{Bus+Metro} & \textbf{Intermodal} \\
		\midrule
		Main: April 2018 $\rightarrow$ April 2019 & $-0.9$ [$-3.3$, $+1.6$] (0.596) & $-4.7$ [$-7.4$, $-1.8$] (0.004) \\
		\addlinespace
		\multicolumn{3}{@{}l}{\emph{Alternative periods and placebo tests}} \\
		May 2018 $\rightarrow$ May 2019 & $+0.0$ [$-2.5$, $+2.6$] (0.991) & $-4.7$ [$-8.1$, $-1.3$] (0.015) \\
		Placebo (pre): April $\rightarrow$ May 2018 & $-1.4$ [$-2.5$, $-0.2$] (0.039) & $-0.0$ [$-1.8$, $+1.7$] (0.972) \\
		Placebo (pre): April 2017 $\rightarrow$ April 2018 & $-0.8$ [$-4.4$, $+2.9$] (0.740) & -- \\
		Placebo (post): April $\rightarrow$ May 2019 & $-0.6$ [$-1.2$, $+0.0$] (0.114) & $-0.4$ [$-1.2$, $+0.4$] (0.363) \\
		\addlinespace
		\multicolumn{3}{@{}l}{\emph{Definition of treated zones}} \\
		E-scooter trips from March 1, 2019 & $-0.3$ [$-3.5$, $+3.0$] (0.905) & $-2.7$ [$-5.9$, $+0.5$] (0.124) \\
		Threshold: $\geq$3 trips per day & $-0.9$ [$-3.3$, $+1.6$] (0.589) & $-3.2$ [$-6.5$, $+0.2$] (0.083) \\
		Threshold: $\geq$10 trips per day & $-0.1$ [$-2.6$, $+2.5$] (0.957) & $-5.2$ [$-7.9$, $-2.4$] (0.003) \\
		Availability-based: all zones of Las Condes & $+0.5$ [$-2.2$, $+3.3$] (0.779) & $-5.0$ [$-8.6$, $-1.1$] (0.028) \\
		\addlinespace
		\multicolumn{3}{@{}l}{\emph{Metro access and Metro Line 3}} \\
		Within 750 m of a metro station & $+2.0$ [$-0.7$, $+4.8$] (0.254) & $-6.0$ [$-8.6$, $-3.3$] (0.001) \\
		No metro station within 750 m & $-4.4$ [$-6.7$, $-1.9$] (0.002) & $-5.4$ [$-9.6$, $-1.0$] (0.015) \\
		Without metro controls & $+0.1$ [$-2.7$, $+3.0$] (0.952) & $-2.9$ [$-5.8$, $+0.1$] (0.068) \\
		Excluding zones near Line 3 & $-0.7$ [$-3.0$, $+1.8$] (0.629) & $-5.2$ [$-7.7$, $-2.6$] (0.002) \\
		\addlinespace
		\multicolumn{3}{@{}l}{\emph{Exclusion buffer and estimator}} \\
		Exclusion buffer: 0 m & $-0.7$ [$-3.1$, $+1.8$] (0.674) & $-4.8$ [$-7.5$, $-2.0$] (0.002) \\
		Exclusion buffer: 2\,000 m & $-0.9$ [$-3.3$, $+1.6$] (0.607) & $-4.7$ [$-7.4$, $-1.8$] (0.004) \\
		Log difference by zone (equal weights) & $-0.9$ (0.482) & $-5.3$ (0.021) \\
		\bottomrule
	\end{tabular}
	\caption{Robustness analysis for public transport trip generation in the Intermediate Region: combined Bus+Metro and intermodal models. Values are percentage changes (Treatment $\times$ Post), with 95\% confidence intervals from zone-clustered standard errors in brackets and randomization-inference $p$-values in parentheses. All models include the controls of the main specification unless indicated. \emph{Placebo (pre)} and \emph{(post)} compare two periods before, or after, the introduction of e-scooters. Baseline: April 9--13, 2018 $\leftrightarrow$ April 8--12, 2019 (Monday--Friday).}
	\label{tab:robustness_generation}
\end{table}

Second, the absence of an effect on combined Bus+Metro demand does not depend on the definition of treated zones, whereas the intermodal effect weakens when zones with little e-scooter activity are counted as treated. Across treatment thresholds from 1 to 15 daily trips, the Bus+Metro effect ranges from $-1.2\%$ to $+0.5\%$, never statistically significant; the intermodal effect ranges from $-2.6\%$ to $-6.8\%$ and is significant at the 5\% level for thresholds of 5 trips or more (Appendix~I). When e-scooter activity is measured from March 1, 2019 instead of the study week (49 and 47 treated zones), the Bus+Metro effect is $-0.3\%$ and $-0.5\%$ and the intermodal effect is $-2.7\%$ ($p=0.12$) and $-4.8\%$ ($p=0.051$). With an availability-based definition, in which all zones of Las Condes are treated regardless of e-scooter use (27 treated zones in the Intermediate Region), the Bus+Metro effect is $+0.5\%$ and $-0.8\%$ and the intermodal effect is $-5.0\%$ ($p=0.028$) and $-5.8\%$ ($p=0.007$) (Appendix~L).

\begin{table}[h]
	\centering
	\scriptsize
	\begin{tabular}{@{}lcc@{}}
		\toprule
		\textbf{Specification} & \textbf{Bus+Metro} & \textbf{Intermodal} \\
		\midrule
		Main: April 2018 $\rightarrow$ April 2019 & $-1.1$ [$-3.8$, $+1.6$] (0.523) & $-5.7$ [$-9.5$, $-1.8$] (0.010) \\
		\addlinespace
		\multicolumn{3}{@{}l}{\emph{Alternative periods and placebo tests}} \\
		May 2018 $\rightarrow$ May 2019 & $-0.3$ [$-3.0$, $+2.5$] (0.895) & $-4.4$ [$-8.8$, $+0.2$] (0.075) \\
		Placebo (pre): April $\rightarrow$ May 2018 & $-1.6$ [$-3.2$, $+0.1$] (0.078) & $-0.6$ [$-3.0$, $+2.0$] (0.717) \\
		Placebo (pre): April 2017 $\rightarrow$ April 2018 & -- & -- \\
		Placebo (post): April $\rightarrow$ May 2019 & $-0.6$ [$-1.5$, $+0.3$] (0.299) & $+0.8$ [$-0.4$, $+2.0$] (0.245) \\
		\addlinespace
		\multicolumn{3}{@{}l}{\emph{Definition of treated zones}} \\
		E-scooter trips from March 1, 2019 & $-0.5$ [$-4.6$, $+3.8$] (0.872) & $-4.8$ [$-9.1$, $-0.3$] (0.051) \\
		Threshold: $\geq$3 trips per day & $+0.2$ [$-4.2$, $+4.8$] (0.939) & $-4.6$ [$-9.1$, $+0.1$] (0.074) \\
		Threshold: $\geq$10 trips per day & $-0.7$ [$-3.6$, $+2.2$] (0.710) & $-6.7$ [$-9.8$, $-3.4$] (0.001) \\
		Availability-based: all zones of Las Condes & $-0.8$ [$-3.7$, $+2.1$] (0.662) & $-5.8$ [$-9.1$, $-2.4$] (0.007) \\
		\addlinespace
		\multicolumn{3}{@{}l}{\emph{Metro access and Metro Line 3}} \\
		Within 750 m of a metro station & $+0.9$ [$-1.8$, $+3.8$] (0.585) & $-2.4$ [$-5.3$, $+0.6$] (0.170) \\
		No metro station within 750 m & $-4.1$ [$-7.8$, $-0.2$] (0.040) & $-10.4$ [$-18.7$, $-1.2$] (0.019) \\
		Without metro controls & $-0.7$ [$-3.7$, $+2.4$] (0.734) & $-4.3$ [$-7.3$, $-1.2$] (0.019) \\
		Excluding zones near Line 3 & $-1.1$ [$-3.8$, $+1.6$] (0.449) & $-6.7$ [$-9.8$, $-3.6$] (0.003) \\
		\addlinespace
		\multicolumn{3}{@{}l}{\emph{Exclusion buffer and estimator}} \\
		Exclusion buffer: 0 m & $-0.9$ [$-3.5$, $+1.9$] (0.627) & $-5.7$ [$-9.4$, $-1.9$] (0.007) \\
		Exclusion buffer: 2\,000 m & $-0.9$ [$-3.6$, $+1.8$] (0.606) & $-5.5$ [$-9.3$, $-1.6$] (0.009) \\
		Log difference by zone (equal weights) & $-4.8$ (0.253) & $-8.5$ (0.113) \\
		\bottomrule
	\end{tabular}
	\caption{Robustness analysis for public transport trip attraction in the Intermediate Region: combined Bus+Metro and intermodal models. Same specifications and notation as Table \ref{tab:robustness_generation}; the placebo between April 2017 and April 2018 is available for trip generation only.}
	\label{tab:robustness_attraction}
\end{table}

Third, the decline in intermodal trips does not reflect differences in metro access or the opening of Metro Line 3. For generation, it is similar among zones within 750~m of a metro station ($-6.0\%$, $p=0.001$) and among zones without one ($-5.4\%$, $p=0.015$), and when zones near Line 3 are excluded ($-5.2\%$, $p=0.002$); for attraction the estimates by metro access are less precise ($-2.4\%$, $p=0.17$; and $-10.4\%$, $p=0.019$). For the combined Bus+Metro model, the effect is $+2.0\%$ and $+0.9\%$ among zones near a metro station and $-4.4\%$ ($p=0.002$) and $-4.1\%$ ($p=0.040$) among zones without one. The latter zones are served by bus, and for them the placebo test between April and May 2018 is also negative ($-2.7\%$, $p<0.001$, for boardings; $-3.1\%$, $p=0.022$, for alightings; Appendix~G), so, as with the bus estimates, we do not interpret this difference as an effect of e-scooters.

Fourth, the estimates are stable across spillover-buffer radii between 0 and 2,000~m ($-0.7\%$ to $-1.1\%$ for Bus+Metro; $-4.7\%$ to $-5.7\%$ for intermodal trips) and under an alternative estimator, the regression of the change in log trips per zone, which weights zones equally ($-0.9\%$ and $-4.8\%$ for Bus+Metro, neither significant; $-5.3\%$, $p=0.021$, and $-8.5\%$, $p=0.11$, for intermodal trips).

Finally, because trip attraction relies on inferred alightings, we examined whether the share of trips without an inferred alighting changed differently in treated zones (Appendix~K). In the Intermediate Region this share went from 22.9\% to 22.5\% among trips starting in treated zones and from 25.9\% to 24.8\% in control zones, a differential change of $+0.6$ percentage points ($p<0.001$), which is small relative to the decline in intermodal trips and does not alter the conclusion for the combined Bus+Metro model.

Taken together, the specifications show that the absence of an effect on combined Bus+Metro demand in the Intermediate Region is stable across periods, treatment definitions and model choices, and that the decline in intermodal trips is consistently negative, although less precise when zones with little e-scooter activity are counted as treated. We therefore treat these two findings as the most robust, and do not interpret the Central and Peripheral estimates, which are based on only two treated zones in each region.

\section{Discussion}
\label{sec:discussion}

Our results show that, at the scale of adoption observed in Santiago, journeys that combine bus and metro declined in the zones where shared e-scooters were used, while the change in total public transport demand is not statistically significant. These results correspond to the Intermediate Region, where the service was concentrated.

\subsection*{Intermodal Trips and the First/Last-Mile Mechanism}

The strict-intermodal model (Tables~\ref{tab:intermodal_generation} and \ref{tab:intermodal_attraction}) restricts the outcome to trips that explicitly combine a bus stage with a metro stage in a single journey. Intermodal generation in the Intermediate Region decreases by $4.7\%$ ($p=0.004$) and intermodal attraction by $5.7\%$ ($p=0.010$), while total boardings and alightings show no statistically significant change. The decline is absent before the introduction of e-scooters, is similar in size with data from another month, and is larger where e-scooter use was more intense. Its direction is consistent with the observation that e-scooters tend to complement rail but compete with buses \citep{jayawardhena2025shared}: the replacement of a short bus stage to or from the metro would reduce bus--metro combinations without reducing the number of public transport trips.

Because bus--metro transfers are observed in the smart-card records, our estimates are not subject to the buffer-sensitivity bias documented by \citet{qian2025big} and \citet{yin2024shared}. The e-scooter and public transport stages of the same journey are not observed, so the relation to first/last-mile use is inferred, and we regard the decline as an association that merits verification with data that link the modes at the level of the journey.

\subsection*{Total Demand and the Role of Scale}

In the region that concentrates e-scooter use, the change in total public transport demand is not statistically significant, and changes larger than about 4\% can be excluded. This agrees with the quasi-experimental studies that measure observed ridership, which report no significant effect or near-zero net effects on bus demand \citep{ziedan2021impacts,ziedan2021complement} or losses confined to the stops exposed to competing trips \citep{luo2021}, and contrasts with US evidence of e-scooter substitution in metro and medium-sized urbanized areas \citep{choi2025effect}.

Scale helps reconcile these findings with studies that classify e-scooter trips by their relation to transit, as potential substitutes or as connections \citep{vinagre2023blind,yan2021spatiotemporal,ahmad2025scooters}. A high share of e-scooter trips related to public transport is compatible with an imperceptible change in transit demand when e-scooter trips are few relative to transit trips: in Santiago, the observed e-scooter trips were equivalent to 0.16\% of boardings in the treated zones. The contradiction across studies likely reflects not only built-environment differences but also how trips are measured: proximity-based classifications are sensitive to analytical choices \citep{qian2025big} and can overstate the transit connections that riders report \citep{yin2024shared}. Substitution and connection rates per e-scooter trip are therefore informative about transit demand only when read together with the relative volume of the two modes.

\subsection*{Mode-Specific Effects and Urban Structure}

Metro boardings in the Intermediate Region show no statistically significant change ($+0.5\%$) once the opening of Metro Line 3 is controlled for, and the estimate for metro alightings is too imprecise to be informative. The negative bus estimates ($-2.9\%$ in boardings and $-3.5\%$ in alightings) cannot be separated from a trend that precedes the introduction of e-scooters, and we do not interpret them.

The effects are estimated with precision in the Intermediate Region, characterized by moderate transit density and intermediate trip distances, where the service was concentrated. In the Central and Peripheral regions the number of treated zones ($N=2$ in each) limits statistical power (see the Supplementary Material, Appendix~H, on small-$N$ inference).

\subsection*{Policy Implications}

At the scale of use observed in Santiago, the introduction of shared e-scooters is compatible with stable overall public transport demand, and its association with public transport appears in journeys that combine bus and metro. Monitoring this relation as adoption grows depends on operators sharing trip data with public-transport authorities, in line with the view of practitioners that cities need to work with operators and use ridership data to plan integration \citep{bruning2025integrating}. If integration with public transport is a policy objective, fare integration has been identified as the single most influential policy lever for combined e-scooter--transit use \citep{aarhaug2026makes}. Finally, because the deployment we study was concentrated in affluent central-eastern areas, regulatory frameworks that set minimum service-area and coverage requirements, instead of leaving siting decisions to operators, can extend micromobility to underserved peripheral zones \citep{bach2023assessing}.

\subsection*{Limitations and Future Research}

\textbf{Measurement and user heterogeneity.} Our analysis operates at the zonal level due to data integration constraints, potentially masking individual-level heterogeneity. In particular, although the e-scooter service was deployed in affluent eastern municipalities, our zone-level data do not identify the socioeconomic profile of individual riders, so a high-income deployment area need not imply high-income riders. Evidence from other settings shows that low-income people do use shared micromobility for utilitarian commuting and last-mile transit trips, though largely where explicit low-income subsidies exist \citep{delbosc2024subsidized}, which Santiago's commercial deployment lacked; we therefore cannot rule out that some riders are workers commuting \emph{into} these job-rich zones who use e-scooters for last-mile egress and access, a pattern aggregate data cannot resolve. Integration of individual-level data across transport modes would enable more precise understanding of modal shift dynamics and support development of adaptive policies that respond to real-time demand patterns across different urban contexts.

\textbf{Buffer-based control-zone definition.} Although we measure intermodal transfers directly from transit records rather than inferring them from proximity to stations, the definition of control zones still depends on a distance buffer around the e-scooter trajectories; the estimates are stable across alternative buffer radii (Supplementary Material, Appendix~I).

\textbf{Generalizability and deployment concentration.} The focus on Santiago limits generalizability, though the systematic spatial approach provides a framework for examining other contexts. The small number of treated zones in the Central and Peripheral clusters ($N=2$ in each) reflects the geographic concentration of the actual e-scooter deployment, which was confined to the eastern mid-upper income areas (Las Condes, Providencia, Vitacura, La Reina) rather than the historical downtown, a concentration consistent with the social and spatial inequities documented for shared micromobility elsewhere \citep{su2024spatial}, in Latin America \citep{desouzaesilva2025micromobility}, and for Santiago specifically \citep{tiznado2021using}. This deployment pattern is an empirical feature of the case study, not an artefact of the clustering procedure: under every alternative clustering approach that yields a usable partition (K-Means with $k=3$--$8$ and SKATER \citep{assuncao2006efficient}; see the Supplementary Material, Appendix~C), the treated zones concentrate in the same contiguous eastern macro-region, while Ward-based partitions are too unbalanced to be informative. Future studies could exploit later e-scooter deployments with broader spatial coverage to obtain estimates for the central and peripheral clusters.

\textbf{Temporal scope.} Short-term effects may not capture longer-term adaptation patterns as users optimize multimodal journeys. Future research should examine temporal variations to understand whether substitution effects help manage peak-hour congestion or create capacity challenges.

\textbf{Treatment defined by use.} In the main analysis, treated zones are those where e-scooters were used, not those where the service was available. Zones with more e-scooter use may differ from other zones in ways that also affect public transport demand. An availability-based definition at the level of the municipality leads to the same conclusions (Appendix~L), but we do not have the exact service-area boundaries. The placebo tests, the use of e-scooter activity measured before the study week, and the comparison with a later month mitigate this concern but do not remove it; moreover, the parallel-trends assumption is supported by a small number of pre-treatment comparisons rather than by a multi-period assessment.

\textbf{Unlinked records and inferred destinations.} Smart-card and e-scooter records cannot be linked to the same traveler or journey, so first/last-mile use is inferred rather than observed. Alightings are also inferred, and about one quarter of trips have no inferred alighting; the inference procedure is the same in both years and the share of unresolved trips changed little in treated and control zones (a differential of 0.6 percentage points), but its accuracy at the zone level cannot be validated with our data.

\textbf{Data availability and operator integration.} A broader obstacle, particularly acute in Chile and Latin America, is that shared-micromobility data are held by private operators that are not integrated into the public-transport fare and data systems. Unlike smart-card transit records, these data are seldom shared for independent evaluation of the service, which constrains the regional evidence base and the reproducibility and comparability of findings; our analysis was feasible only because the operator provided georeferenced trip records directly.

\section{Conclusions}
\label{sec:conclusions}

This study provides quasi-experimental evidence on the effect of shared e-scooters on public transport demand in Santiago. In the Intermediate Region, which concentrates e-scooter use (38 of the 42 treated zones), a strict-intermodal specification shows that trips combining bus and metro declined by about 5\% in e-scooter zones ($-4.7\%$ in generation, $-5.7\%$ in attraction), consistent with a first/last-mile substitution mechanism. By contrast, the change in combined Bus+Metro demand is not statistically significant ($-0.9\%$ in boardings and $-1.1\%$ in alightings, with confidence intervals that exclude changes larger than about 4\%). All models control explicitly for the concurrent opening of Metro Line 3.

These results make three contributions. First, they show that, at an early stage of adoption, the association between shared e-scooters and public transport is concentrated in journeys that combine bus and metro rather than in total demand, given the small scale of e-scooter use relative to public transport, which helps reconcile the substitution and complementarity reported in the literature with the null or small effects found in studies of observed ridership. Second, the automated clustering methodology reduces the arbitrary selection of treatment and control zones and provides a replicable framework for comparative quasi-experimental research, complemented here by zone-level estimation with randomization inference for settings with few treated zones. Third, as one of the first quantifications of e-scooter effects on public-transport demand in Latin America, where prior micromobility evidence has centred on bike- and scooter-sharing equity \citep{tiznado2021using,saud2021inclusive,desouzaesilva2025micromobility}, the study extends the evidence base beyond North American and European contexts toward cities with comparable socioeconomic segregation.

For planners, these findings indicate that, at the scale observed, the introduction of shared e-scooters is compatible with stable overall public transport demand, and that journeys combining bus and metro are where an association with e-scooter activity first appears and deserve monitoring as adoption grows. Meaningful evaluation of micromobility policies, however, ultimately depends on access to operator data.

\subsection*{Acknowledgments}

We thank DTPM and Gonzalo Cort\'es from Scoot Networks for facilitating the data for this study.

\subsection*{CRediT authorship contribution statement}
\textbf{Daniela Opitz}: Conceptualization, Data curation, Formal analysis, Funding acquisition, Investigation, Methodology, Project administration, Resources, Software, Supervision, Visualization, Writing – original draft, Writing – review \&
editing. \textbf{Eduardo Graells-Garrido}: Conceptualization, Data curation, Formal analysis, Funding acquisition, Investigation, Methodology, Project administration, Resources, Software, Supervision, Visualization, Writing – original draft, Writing – review \& editing. \textbf{Jacqueline Arriagada}: Conceptualization, Formal analysis, Investigation, Methodology, Supervision, Writing – original draft, Writing – review \& editing. \textbf{Matilde Rivas}: Conceptualization, Data curation, Formal analysis, Software, Visualization, Writing – original draft, Writing – review \&
editing.  \textbf{Natalia Meza}: Data curation, Software, Visualization, Writing – original draft, Writing – review \& editing.

\subsection*{Data availability}
Public Transport data can be downloaded from the Directory of Metropolitan Transport at \url{https://www.dtpm.cl/index.php/documentos/matrices-de-viaje}, and extended access can be requested through the same website. The authors do not have permission to share e-scooter data. 

\bibliographystyle{elsarticle-harv}
\bibliography{references}

\newpage
\appendix
\section{Time blocks}
\begin{table}[H]
 \scriptsize
\centering
\begin{tabular}{|l|l|}
\hline
\textbf{Time}                & \textbf{Period}         \\ \hline
06:01 - 09:00:59             & Morning Peak      \\ \hline
09:01 - 12:00:59             & Morning Valley    \\ \hline
12:01 - 14:00:59             & Lunch             \\ \hline
14:01 - 17:30:59             & Afternoon Valley  \\ \hline
17:31 - 20:30:59             & Afternoon Peak    \\ \hline
20:31 - 23:00:59             & Night Valley      \\ \hline
23:01 - 06:00:59             & Night             \\ \hline
\end{tabular}
\caption{Time blocks used in the study}
\label{tab:time_blocks}
\end{table}

\newpage
\section{Zone Clustering Variables}
\begin{table}[H]
\singlespacing \scriptsize
\renewcommand{\arraystretch}{0.92}
\setlength{\tabcolsep}{3.5pt}
\centering
\begin{tabularx}{\textwidth}{@{}p{2.4cm} X c p{2.3cm} l r r@{}}
    \toprule
    \textbf{Name} & \textbf{Description} & \textbf{Source} & \textbf{Use} & \textbf{Unit} & \textbf{Mean} & \textbf{Std} \\
    \midrule
    under age population & Percentage of the population younger than 18. & 1 & Zone clustering & \% & 22.50 & 4.78 \\
    working population & Percentage of the population between 18 and 65. & 1 & Zone clustering & \% & 65.60 & 4.44 \\
    retired population & Percentage of the population older than 65. & 1 & Zone clustering & \% & 11.90 & 4.61 \\
    female population & Percentage of the population that is female. & 1 & Zone clustering & \% & 51.50 & 2.11 \\
    avg.\ educational level & Average years of education received by people over 18. & 1 & Zone clustering & years & 11.86 & 1.80 \\
    bus stops & Number of bus stops in the zone. & 2 & Candidate (excluded) & count & 14.09 & 8.11 \\
    bus stop density & Bus stops per square kilometer. & 2 & Zone clustering & stops/km$^{2}$ & 14.15 & 7.90 \\
    bus stops per 1000 people & Bus stops per 1000 inhabitants. & 1, 2 & Candidate (excluded) & stops/person & 2.50 & 4.13 \\
    population density & Inhabitants per square kilometer. & 1 & Zone clustering & person/km$^{2}$ & 7696.17 & 4591.17 \\
    avg.\ number of trips per person & Average number of public transport trips per person. & 3 & Zone clustering / Regression & count & 1.91 & 0.28 \\
    avg.\ time of first trip & Average hour of the first daily public-transport trip. & 3 & Zone clustering & hour & 10.99 & 0.40 \\
    avg.\ trip distance & Average distance of public-transport trips originating in the zone. & 3 & Zone clustering & meters & 9218.33 & 3053.81 \\
    \bottomrule
\end{tabularx}
\caption{Summary of the 12 candidate zone clustering variables (10 retained). Sources: (1) 2017 Census, (2) OpenStreetMap, (3) ADATRAP.}
\label{table:variables}
\end{table}

\FloatBarrier

\begin{table}[H]
\centering
\scriptsize
\begin{tabular}{|l|c|c|}
\hline
\textbf{Name }                 & \textbf{min}       & \textbf{max }    \\  \hline
under age population   & 3.25      & 39.35     \\
working population     & 56.28     & 93.30     \\
retired population     & 1.22      & 24.84     \\
female population      & 10.90     & 56.44     \\
avg. educational level & 8.32      & 15.98     \\
bus stops              & 0.00      & 80.00     \\
bus stop density       & 0.00      & 48.25     \\
bus stops per 1000 people & 0.00   & 77.93     \\
population density     & 31.89     & 34993.67  \\
avg. number of trips per person & 1.00 & 2.71  \\
avg. time of first trip & 9.49     & 13.05     \\
avg. trip distance     & 1627.34   & 20322.69\\ \hline
\end{tabular}
\caption{Minimum and maximum values of each zone clustering variable.}
\label{table:min_max}
\end{table}

\begin{figure}[H]
  \centering
  \includegraphics[width=0.85\linewidth,height=0.78\textheight,keepaspectratio]{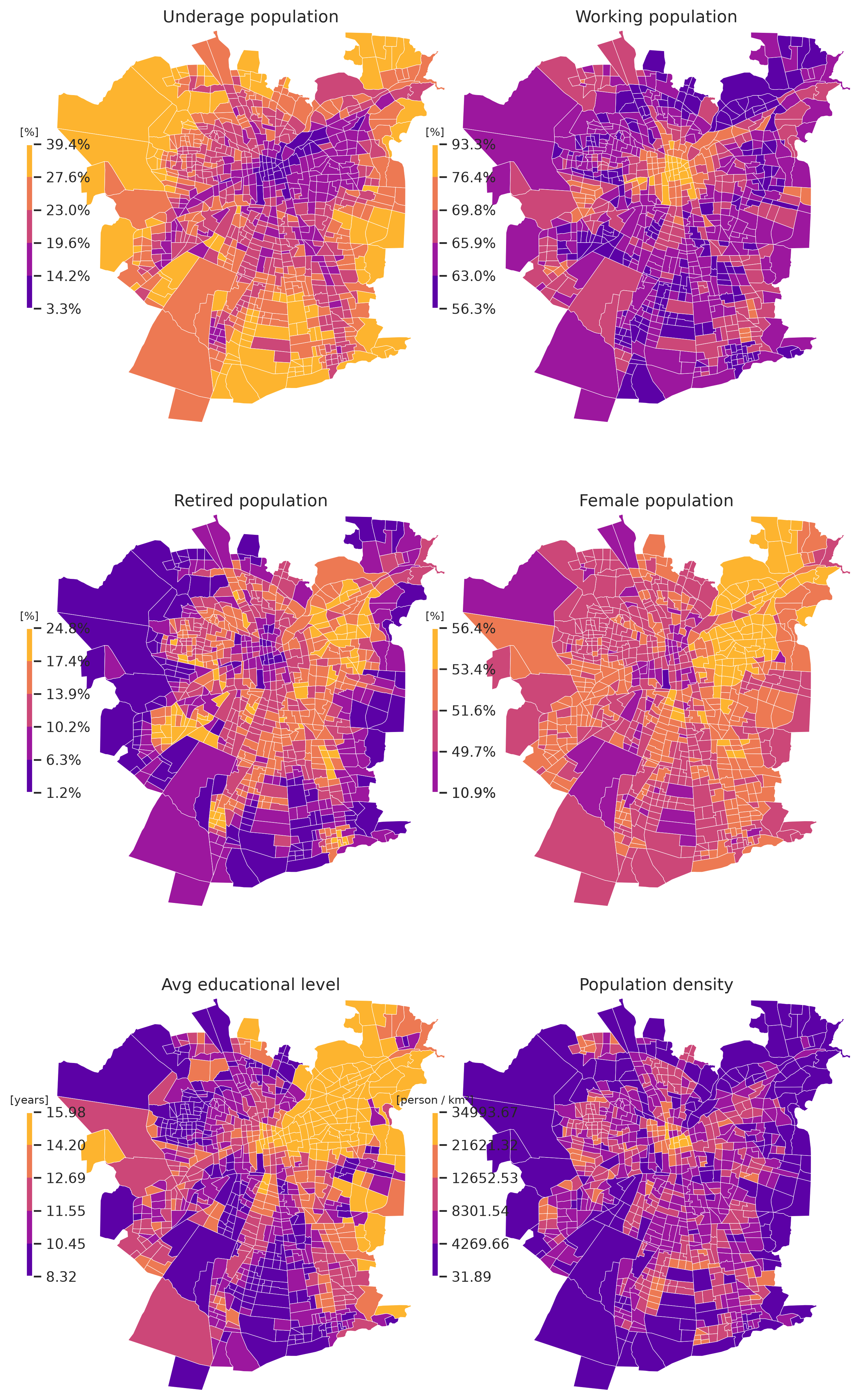}
  \caption{Spatial Distribution of Demographic Characteristics. Maps of underage population, working population, retired population, female population, average educational level, and population density across Traffic Analysis Zones.}
  \label{fig:var1}
\end{figure}

\begin{figure}[H]
  \centering
  \includegraphics[width=0.85\linewidth,height=0.78\textheight,keepaspectratio]{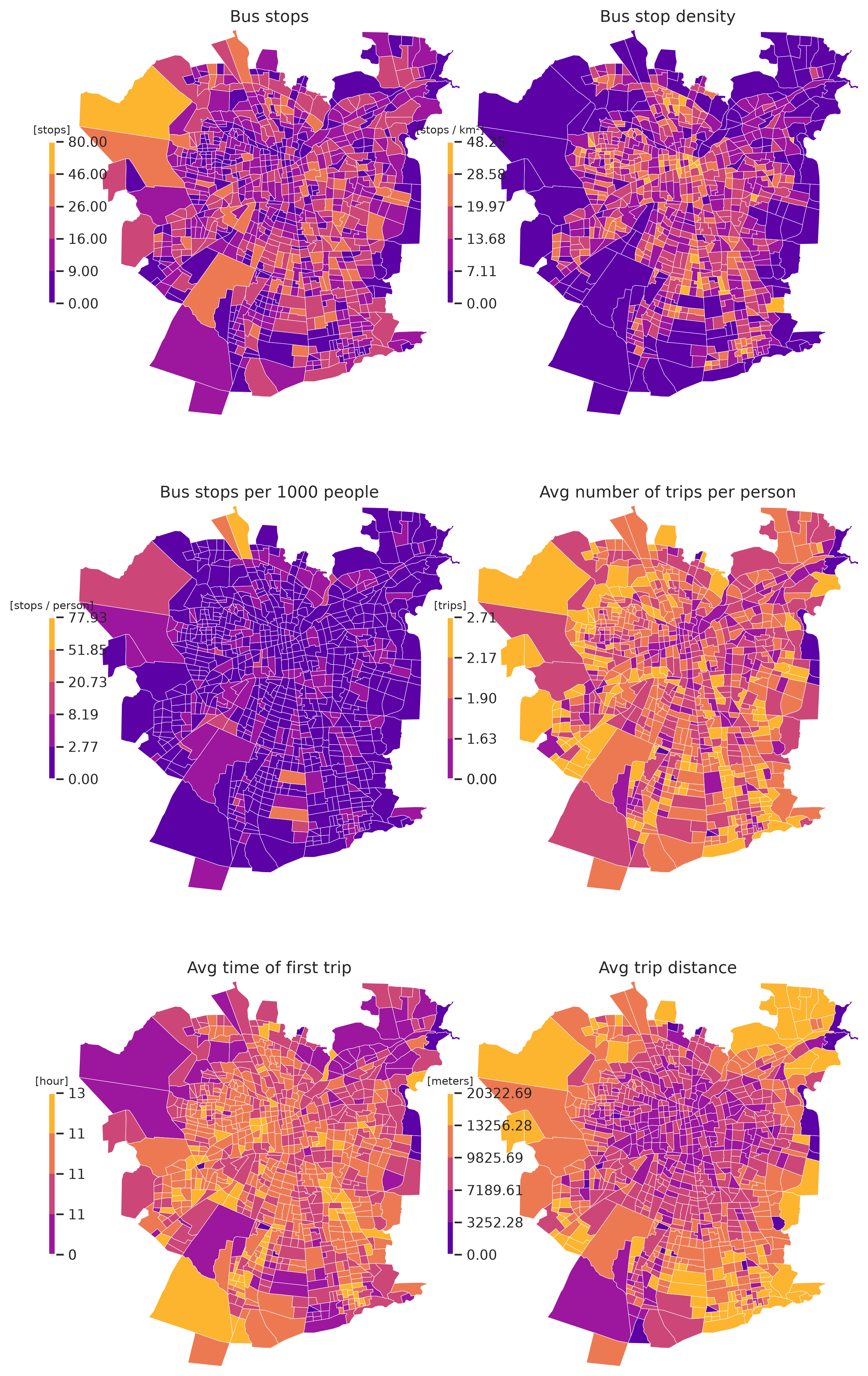}
  \caption{Spatial Distribution of Transportation and Mobility Metrics. Maps of bus stops, bus stop density, bus stops per 1,000 people, average trips per person, average time of the first trip, and average trip distance across Traffic Analysis Zones.}
  \label{fig:var2}
\end{figure}

\subsection*{Distribution of the remaining clustering variables by region}
The main text presents the six variables most relevant to the cluster interpretation. The remaining six candidate clustering variables are shown below as box plots by region. Colours denote regions: \emph{Central} (dark purple), \emph{Peripheral} (magenta), and \emph{Intermediate} (orange).

\begin{figure}[H]
  \centering
  \includegraphics[width=0.95\linewidth]{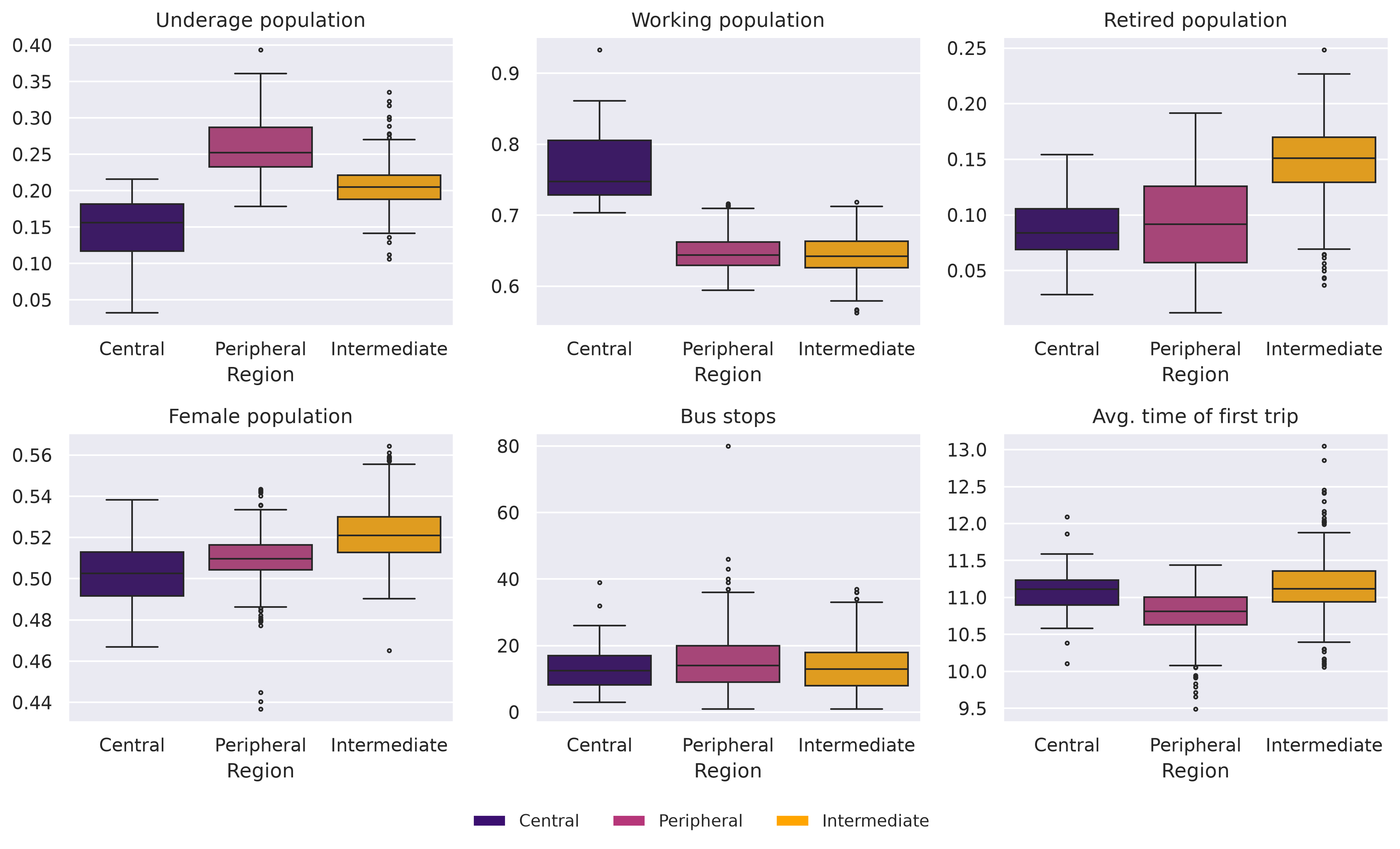}
  \caption{Remaining demographic and transport clustering variables across regions: underage, working, retired and female population, total bus stops, and average time of the first trip.}
  \label{fig:boxplot_rest}
\end{figure}

\FloatBarrier

\section{Clustering Robustness}
\label{sec:clustering_robustness}

The main analysis uses K-Means with $k=3$ over the 10 socioeconomic and accessibility variables retained after the spatial-autocorrelation screening to partition zones into Central, Intermediate, and Peripheral. We assess whether this partition, and the concentration of treated zones in one region, are sensitive to this choice by re-running the clustering with four alternative methods: hierarchical agglomerative (Ward), regionalisation with Queen-contiguity constraint, regionalisation with KNN connectivity, and SKATER (Assun\c{c}\~ao et al., 2006) with a minimum-cluster-size floor.

Three observations emerge. First, K-Means and SKATER both place the bulk of treated zones (49.2\% concentration in the SKATER 63-zone cluster) in a contiguous eastern macro-region corresponding to Las Condes, Providencia, Vitacura, and La Reina. This is a feature of the empirical e-scooter deployment in Santiago, not of any particular clustering algorithm. Second, Ward+Queen and Ward+KNN with $k=3$ produce highly unbalanced clusters (a single large cluster of $\approx 790$ zones plus 2--3 outlier clusters), which is not suitable for stratified DiD; this is consistent with the limited number of underlying socioeconomic ``regimes'' in Santiago's spatial structure. Third, increasing $k$ in K-Means progressively concentrates the treated zones into a smaller and more homogeneous cluster (27.9\% treated at $k=8$).

\begin{figure}[H]
  \centering
  \includegraphics[width=0.95\linewidth]{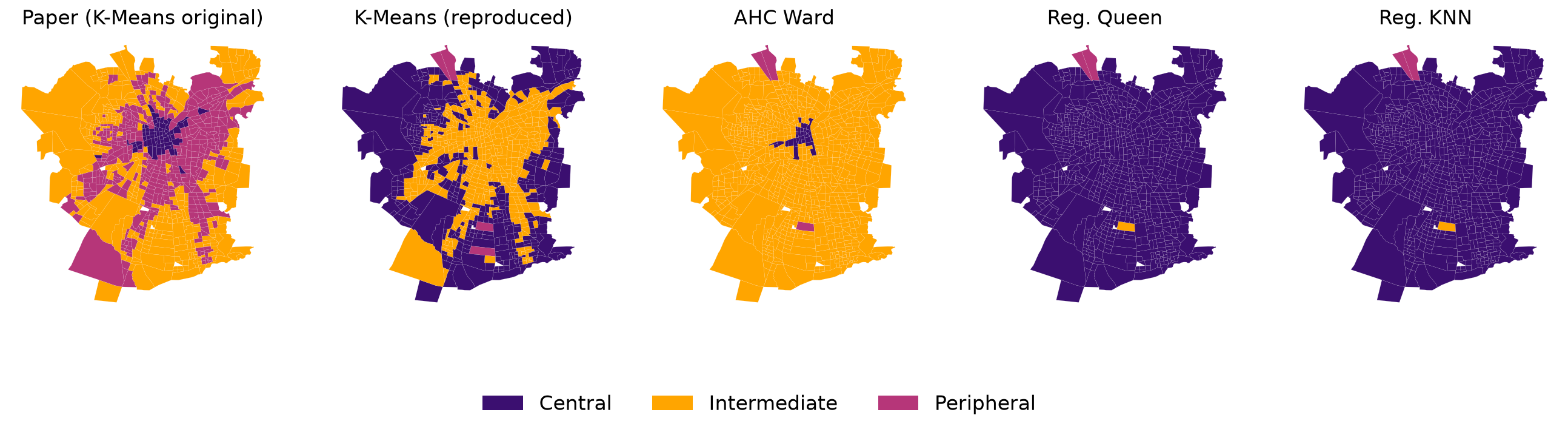}
  \caption{Spatial partitions produced by K-Means (paper), K-Means reproduced, AHC Ward, Reg. Queen, and Reg. KNN, all with $k=3$. Treated zones (concentration of e-scooter activity) consistently fall in the same eastern macro-region; the Ward-based partitions collapse into a single dominant cluster and are shown for completeness.}
  \label{fig:clustering_methods}
\end{figure}

\begin{figure}[H]
  \centering
  \includegraphics[width=0.75\linewidth]{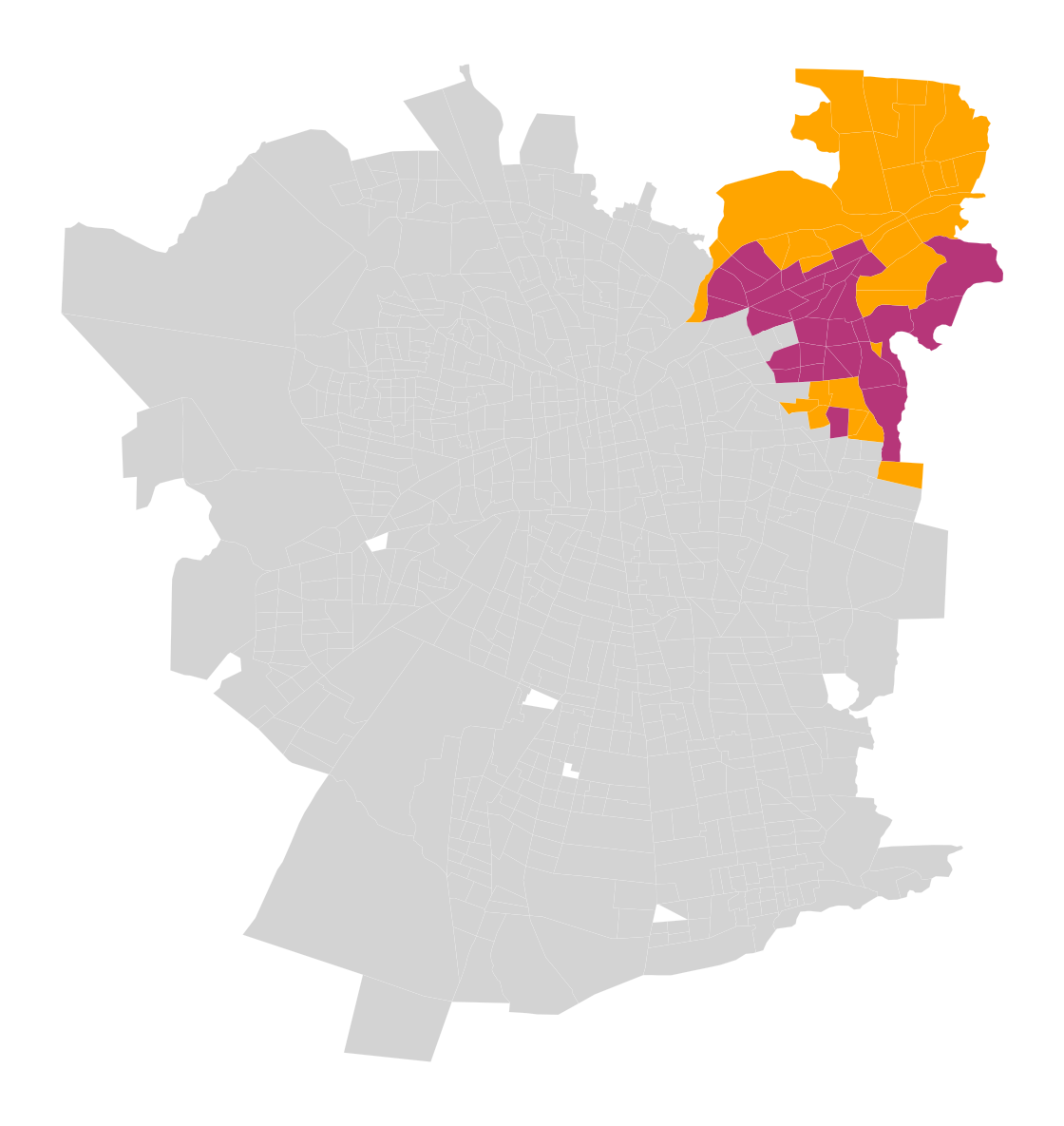}
  \caption{SKATER cluster (orange) containing the bulk of treated zones (magenta), corresponding to the eastern macro-region of Santiago: Las Condes, Providencia, Vitacura, and La Reina. 31 of the 63 zones in this SKATER cluster are treated (49.2\%).}
  \label{fig:clustering_skater}
\end{figure}

\clearpage

\clearpage
\section{Results of the Models for Evaluating Generation of Public Transport Trips}
\label{sec:full_generation}

Tables~\ref{tab:full_generation} and~\ref{tab:full_attraction} report the complete output of the difference-in-differences models for every model and region. All estimates cited in the main text and in the following appendices are produced by a single run of the same code on the same data, and Tables~2, 4, 5 and~6 of the main text are subsets of these two tables.

For each cell the tables give the number of treated zones and of zones in the model; the trips of the 2018 period in treated and control zones (thousands, over the five working days); the Treatment $\times$ Post coefficient $\hat\beta_3$ with its zone-clustered standard error; the effect $\exp(\hat\beta_3)-1$ and its 95\% confidence interval; the cluster-robust $p$-value; the randomization-inference $p$-value with the number of assignments used (complete enumeration or 5,000 random draws); and, as an alternative estimator, the coefficient of an ordinary least squares regression of the change in log trips per zone, which weights zones equally, with its randomization-inference $p$-value. The Metro models include only the zones with metro trips.

In the Central and Peripheral regions two zones are treated. For these cells the cluster-robust standard errors, $p$-values and intervals are shown for completeness, but they are not reliable with so few treated clusters; inference rests on the randomization $p$-values (Appendix~H).

\begin{table}[H]
\centering
\singlespacing
\scriptsize
\setlength{\tabcolsep}{3.5pt}
\resizebox{\textwidth}{!}{\begin{tabular}{@{}llrrrrrrrrlr@{}}
\toprule
Outcome & Region & \makecell{Treated/\\zones} & \makecell{Pre trips,\\treated} & \makecell{Pre trips,\\control} & $\hat\beta$ & SE & \makecell{Effect\\(\%)} & 95\% CI & \makecell{$p$\\(robust)} & \makecell[l]{$p$ (RI);\\assignments} & \makecell{Log diff.\\(\%); $p$ (RI)} \\
\midrule
Bus+Metro & Central & 2/58 & 136 & 5,888 & -0.0588 & 0.0135 & $-5.7$ & [$-8.2$, $-3.2$] & $<$0.001 & 0.023; exact (1,653) & $-11.6$; 0.294 \\
 & Intermediate & 38/361 & 2,003 & 6,573 & -0.0085 & 0.0127 & $-0.9$ & [$-3.3$, $+1.6$] & 0.501 & 0.596; 5,000 random & $-0.9$; 0.482 \\
 & Peripheral & 2/320 & 44 & 4,570 & +0.0111 & 0.0115 & $+1.1$ & [$-1.1$, $+3.4$] & 0.336 & 0.694; 5,000 random & $-1.0$; 0.876 \\
\addlinespace
Bus & Central & 2/58 & 136 & 1,717 & +0.0103 & 0.0414 & $+1.0$ & [$-6.8$, $+9.6$] & 0.803 & 0.882; exact (1,653) & $-6.4$; 0.577 \\
 & Intermediate & 38/361 & 709 & 3,656 & -0.0291 & 0.0106 & $-2.9$ & [$-4.9$, $-0.8$] & 0.006 & 0.009; 5,000 random & $-1.6$; 0.118 \\
 & Peripheral & 2/320 & 44 & 3,468 & +0.0170 & 0.0113 & $+1.7$ & [$-0.5$, $+4.0$] & 0.131 & 0.526; 5,000 random & $-0.7$; 0.909 \\
\addlinespace
Metro & Central & 0/25 & -- & -- & -- & -- & -- & -- & -- & -- & -- \\
 & Intermediate & 8/61 & 1,295 & 2,970 & +0.0049 & 0.0160 & $+0.5$ & [$-2.6$, $+3.7$] & 0.761 & 0.865; 4,997 random & $-3.2$; 0.256 \\
 & Peripheral & 0/19 & -- & -- & -- & -- & -- & -- & -- & -- & -- \\
\addlinespace
Intermodal & Central & 2/58 & 6 & 796 & +0.0418 & 0.0724 & $+4.3$ & [$-9.5$, $+20.2$] & 0.563 & 0.780; exact (1,653) & $-4.8$; 0.726 \\
 & Intermediate & 38/359 & 398 & 1,325 & -0.0479 & 0.0150 & $-4.7$ & [$-7.4$, $-1.8$] & 0.001 & 0.004; 5,000 random & $-5.3$; 0.021 \\
 & Peripheral & 2/320 & 29 & 1,632 & -0.0117 & 0.0098 & $-1.2$ & [$-3.0$, $+0.7$] & 0.230 & 0.621; 5,000 random & $-2.6$; 0.389 \\
\bottomrule
\end{tabular}}
\caption{Complete difference-in-differences estimates for trip generation (boardings): April 9--13, 2018 versus April 8--12, 2019. \emph{Bus+Metro}: all trips whose first stage is bus or metro; \emph{Intermodal}: trips combining bus and metro within a single journey. RI: randomization inference.}
\label{tab:full_generation}
\end{table}

\section{Results of the Models for Evaluating Attraction of Public Transport Trips}
\label{sec:full_attraction}

\begin{table}[H]
\centering
\singlespacing
\scriptsize
\setlength{\tabcolsep}{3.5pt}
\resizebox{\textwidth}{!}{\begin{tabular}{@{}llrrrrrrrrlr@{}}
\toprule
Outcome & Region & \makecell{Treated/\\zones} & \makecell{Pre trips,\\treated} & \makecell{Pre trips,\\control} & $\hat\beta$ & SE & \makecell{Effect\\(\%)} & 95\% CI & \makecell{$p$\\(robust)} & \makecell[l]{$p$ (RI);\\assignments} & \makecell{Log diff.\\(\%); $p$ (RI)} \\
\midrule
Bus+Metro & Central & 2/58 & 92 & 6,312 & -0.0365 & 0.0109 & $-3.6$ & [$-5.6$, $-1.5$] & $<$0.001 & 0.082; exact (1,653) & $-7.1$; 0.456 \\
 & Intermediate & 38/361 & 2,126 & 6,754 & -0.0113 & 0.0139 & $-1.1$ & [$-3.8$, $+1.6$] & 0.417 & 0.523; 5,000 random & $-4.8$; 0.253 \\
 & Peripheral & 2/320 & 25 & 3,913 & +0.1144 & 0.0275 & $+12.1$ & [$+6.2$, $+18.3$] & $<$0.001 & 0.233; 5,000 random & $+2.9$; 0.863 \\
\addlinespace
Bus & Central & 2/58 & 92 & 1,595 & +0.0169 & 0.0266 & $+1.7$ & [$-3.5$, $+7.1$] & 0.524 & 0.711; exact (1,653) & $-4.1$; 0.708 \\
 & Intermediate & 38/361 & 670 & 3,489 & -0.0354 & 0.0161 & $-3.5$ & [$-6.5$, $-0.4$] & 0.028 & 0.032; 5,000 random & $-5.8$; 0.160 \\
 & Peripheral & 2/320 & 25 & 2,783 & +0.1187 & 0.0271 & $+12.6$ & [$+6.8$, $+18.8$] & $<$0.001 & 0.216; 4,999 random & $+3.3$; 0.840 \\
\addlinespace
Metro & Central & 0/28 & -- & -- & -- & -- & -- & -- & -- & -- & -- \\
 & Intermediate & 20/105 & 1,455 & 3,265 & -0.1147 & 0.1691 & $-10.8$ & [$-36.0$, $+24.2$] & 0.498 & 0.559; 5,000 random & $+14.8$; 0.313 \\
 & Peripheral & 1/52 & $<$1 & 1,130 & +0.0968 & -- & $+10.2$ & -- & -- & 0.863; exact (52) & $-3.4$; 0.500 \\
\addlinespace
Intermodal & Central & 2/58 & 3 & 1,235 & +0.0159 & 0.0663 & $+1.6$ & [$-10.8$, $+15.7$] & 0.810 & 0.811; exact (1,653) & $+9.5$; 0.809 \\
 & Intermediate & 38/360 & 531 & 1,331 & -0.0591 & 0.0208 & $-5.7$ & [$-9.5$, $-1.8$] & 0.004 & 0.010; 5,000 random & $-8.5$; 0.113 \\
 & Peripheral & 2/318 & 11 & 1,031 & +0.0598 & 0.0302 & $+6.2$ & [$+0.1$, $+12.6$] & 0.048 & 0.478; 5,000 random & $-4.5$; 0.789 \\
\bottomrule
\end{tabular}}
\caption{Complete difference-in-differences estimates for trip attraction (alightings): April 9--13, 2018 versus April 8--12, 2019. Columns as in Table~\ref{tab:full_generation}. One treated zone of the Peripheral region has metro alightings, with a negligible number of trips; no standard error can be computed for that cell.}
\label{tab:full_attraction}
\end{table}

\clearpage
\section{Placebo Tests and Alternative Periods}
\label{sec:parallel_trends}

To address the concern that the two-period DiD design does not permit a direct test of the parallel-trends assumption, we refitted the same zone-level DiD model to alternative period pairings, using the working days of each window. The placebo tests compare periods with the same e-scooter status: April 2018 vs. May 2018 and April 2017 vs. April 2018 (the latter for trip generation, with an explicit \texttt{L6\_post} control, since Metro Line 6 opened between them in November 2017), both pre-treatment, and April 2019 vs. May 2019, both post-treatment. A statistically significant Treatment $\times$ Post coefficient in these comparisons would indicate divergence between treated and control zones unrelated to the e-scooter introduction. We also re-estimated the main comparison with May 2018 vs. May 2019. The treated zones are those of the main analysis in all comparisons.

\begin{figure}[H]
  \centering
  \includegraphics[width=0.95\linewidth]{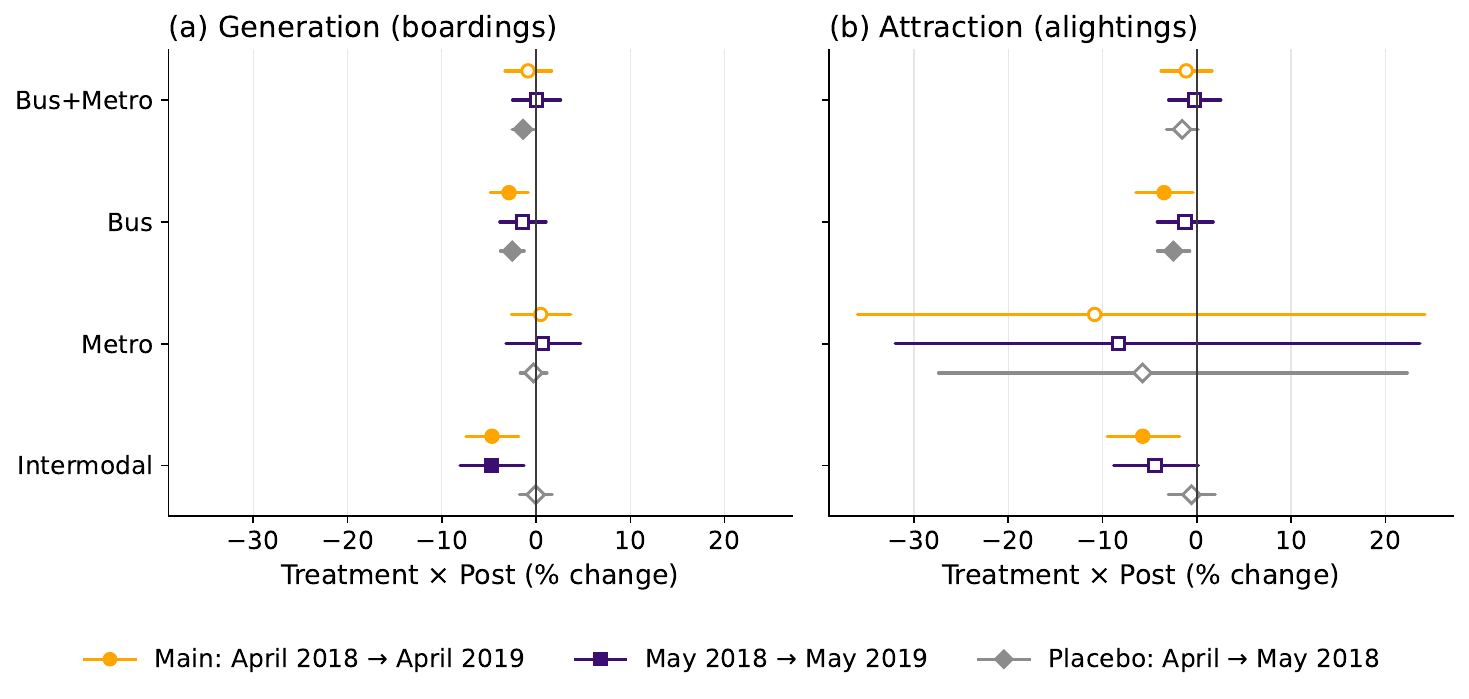}
  \caption{Treatment $\times$ Post estimates (95\% confidence intervals) for the Intermediate Region across three period pairings: main comparison (April 2018 vs. April 2019), May 2018 vs. May 2019, and pre-treatment placebo (April 2018 vs. May 2018). Filled markers indicate a randomization-inference $p$-value below 0.05.}
  \label{fig:forest_validation}
\end{figure}

Tables~\ref{tab:periods} and~\ref{tab:periods_modes} report the results for the Intermediate Region, and Figure~\ref{fig:forest_validation} summarises them. For the intermodal model, the placebo estimates are close to zero and not statistically significant in any comparison. For the Metro model they are not statistically significant either, although they are imprecise (for example, $-16.1\%$, with a randomization-inference $p$-value of 0.12, for boardings between April 2017 and April 2018). For the Bus model, the placebo estimates are negative and significant between April and May 2018 ($-2.5\%$ for boardings and for alightings) and between April and May 2019 for boardings ($-1.3\%$), which indicates a differential trend in bus trips in the treated zones that is unrelated to the e-scooter introduction; Bus-specific estimates are therefore not interpreted in the main text. This trend also appears, attenuated, in the combined Bus+Metro model between April and May 2018 ($-1.4\%$, $p=0.039$, for boardings; $-1.6\%$, $p=0.078$, for alightings). These tests provide supporting evidence for the parallel-trends assumption for the intermodal and Metro models; they do not establish it, and the available windows do not allow a multi-period event-study assessment. The normalised trajectories are presented in the main text (Figure~9).

\begin{table}[H]
\centering
\singlespacing
\scriptsize
\setlength{\tabcolsep}{3.5pt}
\begin{tabular}{@{}>{\raggedright\arraybackslash}p{3.1cm}cccc@{}}
\toprule
 & \multicolumn{2}{c}{Bus+Metro} & \multicolumn{2}{c}{Intermodal} \\
\cmidrule(lr){2-3}\cmidrule(lr){4-5}
 & Generation & Attraction & Generation & Attraction \\
\midrule
Main: April 2018 $\rightarrow$ April 2019 & \makecell{$-0.9$ [$-3.3$, $+1.6$] \\ $p$ = 0.596; 38/361} & \makecell{$-1.1$ [$-3.8$, $+1.6$] \\ $p$ = 0.523; 38/361} & \makecell{$-4.7$ [$-7.4$, $-1.8$] \\ $p$ = 0.004; 38/359} & \makecell{$-5.7$ [$-9.5$, $-1.8$] \\ $p$ = 0.010; 38/360} \\
May 2018 $\rightarrow$ May 2019 & \makecell{$+0.0$ [$-2.5$, $+2.6$] \\ $p$ = 0.991; 38/361} & \makecell{$-0.3$ [$-3.0$, $+2.5$] \\ $p$ = 0.895; 38/361} & \makecell{$-4.7$ [$-8.1$, $-1.3$] \\ $p$ = 0.015; 38/360} & \makecell{$-4.4$ [$-8.8$, $+0.2$] \\ $p$ = 0.075; 38/361} \\
Placebo (pre): April $\rightarrow$ May 2018 & \makecell{$-1.4$ [$-2.5$, $-0.2$] \\ $p$ = 0.039; 38/361} & \makecell{$-1.6$ [$-3.2$, $+0.1$] \\ $p$ = 0.078; 38/361} & \makecell{$-0.0$ [$-1.8$, $+1.7$] \\ $p$ = 0.972; 38/360} & \makecell{$-0.6$ [$-3.0$, $+2.0$] \\ $p$ = 0.717; 38/360} \\
Placebo (pre): April 2017 $\rightarrow$ April 2018 & \makecell{$-0.8$ [$-4.4$, $+2.9$] \\ $p$ = 0.740; 38/361} & -- & -- & -- \\
Placebo (post): April $\rightarrow$ May 2019 & \makecell{$-0.6$ [$-1.2$, $+0.0$] \\ $p$ = 0.114; 38/361} & \makecell{$-0.6$ [$-1.5$, $+0.3$] \\ $p$ = 0.299; 38/361} & \makecell{$-0.4$ [$-1.2$, $+0.4$] \\ $p$ = 0.363; 38/360} & \makecell{$+0.8$ [$-0.4$, $+2.0$] \\ $p$ = 0.245; 38/361} \\
\bottomrule
\end{tabular}
\caption{Alternative periods and placebo tests for the Intermediate Region: Bus+Metro and intermodal models. Each cell reports the effect in percent with its 95\% confidence interval and, below, the randomization-inference $p$-value and the number of treated zones and zones in the model. \emph{Placebo (pre)} and \emph{(post)}: both periods before, or after, the introduction of e-scooters.}
\label{tab:periods}
\end{table}

\begin{table}[H]
\centering
\singlespacing
\scriptsize
\setlength{\tabcolsep}{3.5pt}
\begin{tabular}{@{}>{\raggedright\arraybackslash}p{3.1cm}cccc@{}}
\toprule
 & \multicolumn{2}{c}{Bus} & \multicolumn{2}{c}{Metro} \\
\cmidrule(lr){2-3}\cmidrule(lr){4-5}
 & Generation & Attraction & Generation & Attraction \\
\midrule
Main: April 2018 $\rightarrow$ April 2019 & \makecell{$-2.9$ [$-4.9$, $-0.8$] \\ $p$ = 0.009; 38/361} & \makecell{$-3.5$ [$-6.5$, $-0.4$] \\ $p$ = 0.032; 38/361} & \makecell{$+0.5$ [$-2.6$, $+3.7$] \\ $p$ = 0.865; 8/61} & \makecell{$-10.8$ [$-36.0$, $+24.2$] \\ $p$ = 0.559; 20/105} \\
May 2018 $\rightarrow$ May 2019 & \makecell{$-1.4$ [$-3.8$, $+1.1$] \\ $p$ = 0.338; 38/361} & \makecell{$-1.3$ [$-4.2$, $+1.7$] \\ $p$ = 0.502; 38/361} & \makecell{$+0.7$ [$-3.2$, $+4.7$] \\ $p$ = 0.858; 8/61} & \makecell{$-8.3$ [$-32.0$, $+23.7$] \\ $p$ = 0.609; 16/85} \\
Placebo (pre): April $\rightarrow$ May 2018 & \makecell{$-2.5$ [$-3.8$, $-1.3$] \\ $p<$0.001; 38/361} & \makecell{$-2.5$ [$-4.2$, $-0.8$] \\ $p$ = 0.002; 38/361} & \makecell{$-0.3$ [$-1.7$, $+1.1$] \\ $p$ = 0.768; 8/58} & \makecell{$-5.8$ [$-27.4$, $+22.3$] \\ $p$ = 0.666; 17/93} \\
Placebo (pre): April 2017 $\rightarrow$ April 2018 & \makecell{$-0.9$ [$-3.7$, $+2.1$] \\ $p$ = 0.632; 38/361} & -- & \makecell{$-16.1$ [$-29.4$, $-0.4$] \\ $p$ = 0.123; 8/58} & -- \\
Placebo (post): April $\rightarrow$ May 2019 & \makecell{$-1.3$ [$-2.1$, $-0.4$] \\ $p$ = 0.009; 38/361} & \makecell{$-0.2$ [$-1.1$, $+0.7$] \\ $p$ = 0.705; 38/361} & \makecell{$-0.2$ [$-1.0$, $+0.5$] \\ $p$ = 0.614; 8/61} & \makecell{$-9.0$ [$-31.2$, $+20.5$] \\ $p$ = 0.526; 18/95} \\
\bottomrule
\end{tabular}
\caption{Alternative periods and placebo tests for the Intermediate Region: Bus and Metro models. Cells as in Table~\ref{tab:periods}.}
\label{tab:periods_modes}
\end{table}

\section{Metro Line 3 Confounding Robustness}
\label{sec:l3_robustness}

Metro Line 3 (Santiago) opened on 22 January 2019, between the pre- and post-treatment periods of our DiD design. To assess whether the estimates are confounded with this infrastructure change, we re-estimated the Bus+Metro and intermodal models for the Intermediate Region under three specifications:
\begin{enumerate}
  \item \textbf{Spec 0}: without metro controls.
  \item \textbf{Spec 1 (main specification)}: with the controls of the main specification for Metro Line~3 (including the time-variant \texttt{L3\_post} indicator, equal to 1 for zones within 750\,m of an L3 station in the post period) and for metro access.
  \item \textbf{Spec 2}: excludes from the sample any zone within 750\,m of an L3 station.
\end{enumerate}

Only one of the 38 treated zones of the Intermediate cluster lies within the 750\,m L3 buffer. Direct geographic confounding is therefore limited. The Bus+Metro estimates remain close to zero across Spec 0--2, and the intermodal decline is between $-2.9\%$ and $-6.7\%$ (Table~\ref{tab:metro}).

The table also reports the estimates within groups of zones with comparable metro access in 2018: zones within 750\,m of a metro station, zones within 750\,m of a Line 1 station, and zones without a station within 750\,m. The decline in intermodal generation is similar in the three groups ($-6.0\%$, $-6.0\%$ and $-5.4\%$). For the Bus+Metro model, the estimate is negative among zones without a metro station nearby ($-4.4\%$ for boardings and $-4.1\%$ for alightings). The last block of the table shows that this difference is already present in the placebo test between April and May 2018 ($-2.7\%$ for boardings and $-3.1\%$ for alightings), before the introduction of e-scooters, and is smaller and not statistically significant between May 2018 and May 2019; it is therefore not interpreted as an effect. For intermodal trips the placebo estimates are close to zero in both groups of zones.

\begin{table}[H]
\centering
\singlespacing
\scriptsize
\setlength{\tabcolsep}{3.5pt}
\begin{tabular}{@{}>{\raggedright\arraybackslash}p{3.1cm}cccc@{}}
\toprule
 & \multicolumn{2}{c}{Bus+Metro} & \multicolumn{2}{c}{Intermodal} \\
\cmidrule(lr){2-3}\cmidrule(lr){4-5}
 & Generation & Attraction & Generation & Attraction \\
\midrule
\multicolumn{5}{@{}l}{\emph{Metro Line 3}} \\
Without metro controls & \makecell{$+0.1$ [$-2.7$, $+3.0$] \\ $p$ = 0.952; 38/361} & \makecell{$-0.7$ [$-3.7$, $+2.4$] \\ $p$ = 0.734; 38/361} & \makecell{$-2.9$ [$-5.8$, $+0.1$] \\ $p$ = 0.068; 38/359} & \makecell{$-4.3$ [$-7.3$, $-1.2$] \\ $p$ = 0.019; 38/360} \\
Excluding zones near Line 3 & \makecell{$-0.7$ [$-3.0$, $+1.8$] \\ $p$ = 0.629; 37/333} & \makecell{$-1.1$ [$-3.8$, $+1.6$] \\ $p$ = 0.449; 37/333} & \makecell{$-5.2$ [$-7.7$, $-2.6$] \\ $p$ = 0.002; 37/331} & \makecell{$-6.7$ [$-9.8$, $-3.6$] \\ $p$ = 0.003; 37/332} \\
\addlinespace
\multicolumn{5}{@{}l}{\emph{Metro access}} \\
Within 750 m of a metro station & \makecell{$+2.0$ [$-0.7$, $+4.8$] \\ $p$ = 0.254; 18/212} & \makecell{$+0.9$ [$-1.8$, $+3.8$] \\ $p$ = 0.585; 18/212} & \makecell{$-6.0$ [$-8.6$, $-3.3$] \\ $p$ = 0.001; 18/212} & \makecell{$-2.4$ [$-5.3$, $+0.6$] \\ $p$ = 0.170; 18/212} \\
Within 750 m of Metro Line 1 & \makecell{$-0.2$ [$-3.4$, $+3.1$] \\ $p$ = 0.900; 15/36} & \makecell{$-0.8$ [$-4.2$, $+2.8$] \\ $p$ = 0.703; 15/36} & \makecell{$-6.0$ [$-9.6$, $-2.2$] \\ $p$ = 0.005; 15/36} & \makecell{$-3.2$ [$-7.1$, $+0.9$] \\ $p$ = 0.174; 15/36} \\
No metro station within 750 m & \makecell{$-4.4$ [$-6.7$, $-1.9$] \\ $p$ = 0.002; 20/149} & \makecell{$-4.1$ [$-7.8$, $-0.2$] \\ $p$ = 0.040; 20/149} & \makecell{$-5.4$ [$-9.6$, $-1.0$] \\ $p$ = 0.015; 20/147} & \makecell{$-10.4$ [$-18.7$, $-1.2$] \\ $p$ = 0.019; 20/148} \\
\addlinespace
\multicolumn{5}{@{}l}{\emph{Metro access: placebo and May comparisons}} \\
Placebo (pre), within 750 m of metro & \makecell{$-0.4$ [$-1.7$, $+0.9$] \\ $p$ = 0.682; 18/212} & \makecell{$-0.6$ [$-2.1$, $+0.8$] \\ $p$ = 0.489; 18/212} & \makecell{$+0.8$ [$-1.9$, $+3.5$] \\ $p$ = 0.661; 18/212} & \makecell{$+0.6$ [$-1.5$, $+2.7$] \\ $p$ = 0.670; 18/212} \\
Placebo (pre), no metro within 750 m & \makecell{$-2.7$ [$-4.1$, $-1.3$] \\ $p<$0.001; 20/149} & \makecell{$-3.1$ [$-5.9$, $-0.3$] \\ $p$ = 0.022; 20/149} & \makecell{$-0.5$ [$-2.4$, $+1.5$] \\ $p$ = 0.673; 20/148} & \makecell{$-2.3$ [$-6.3$, $+1.8$] \\ $p$ = 0.338; 20/148} \\
May $\rightarrow$ May, within 750 m of metro & \makecell{$+1.7$ [$-1.4$, $+5.0$] \\ $p$ = 0.428; 18/212} & \makecell{$+0.8$ [$-2.3$, $+3.9$] \\ $p$ = 0.719; 18/212} & \makecell{$-6.7$ [$-9.9$, $-3.3$] \\ $p<$0.001; 18/212} & \makecell{$-3.7$ [$-6.8$, $-0.5$] \\ $p$ = 0.069; 18/212} \\
May $\rightarrow$ May, no metro within 750 m & \makecell{$-1.7$ [$-4.6$, $+1.4$] \\ $p$ = 0.399; 20/149} & \makecell{$-1.5$ [$-5.7$, $+3.0$] \\ $p$ = 0.713; 20/149} & \makecell{$-5.3$ [$-9.9$, $-0.5$] \\ $p$ = 0.042; 20/148} & \makecell{$-5.3$ [$-14.5$, $+5.0$] \\ $p$ = 0.394; 20/149} \\
\bottomrule
\end{tabular}
\caption{Metro Line 3 and metro access, Intermediate Region. Cells as in Table~\ref{tab:periods}.}
\label{tab:metro}
\end{table}

\section{Small-$N$ Inference for the Central and Peripheral Regions}
\label{sec:small_n_central}

The Central and Peripheral clusters contain only $N=2$ treated zones each after applying the 1\,440\,m spillover buffer. With so few treated clusters, cluster-robust standard errors understate the uncertainty of the estimates. We therefore base inference on randomization inference: treatment is reassigned among the zones of the cluster, keeping the number of treated zones fixed, the model is refitted for each assignment, and the observed $t$-statistic is compared with the resulting distribution. In the Central cluster all 1,653 possible assignments of two treated zones are enumerated; in the other cells we use 5,000 random assignments.

The difference between the two approaches can be substantial in these clusters. For the Bus+Metro attraction model in the Peripheral cluster, the estimate is $+12.1\%$ with a cluster-robust 95\% interval of $+6.2\%$ to $+18.3\%$, but the randomization-inference $p$-value is 0.23. In the Central cluster, the Bus+Metro estimates are $-5.7\%$ for generation (randomization-inference $p=0.023$) and $-3.6\%$ for attraction ($p=0.08$). Even when the randomization $p$-value is small, an estimate based on two treated zones reflects the behavior of those two zones: between April 2017 and April 2018, before the introduction of e-scooters, their boardings had already changed by $-19.0\%$ relative to the control zones. In the Intermediate cluster, with 38 treated zones, both approaches lead to the same conclusions (Tables~\ref{tab:full_generation} and~\ref{tab:full_attraction}). For this reason, the main text reports confidence intervals only for cells with at least five treated zones and does not interpret the Central and Peripheral estimates.

\section{Threshold and Buffer Sensitivity}
\label{sec:threshold_buffer}

The main analysis defines a zone as treated if it received at least 5 e-scooter trips per day during the study week, and excludes control zones that fall within a 1\,440\,m spillover buffer of the e-scooter trajectories. We assess sensitivity to both choices, the daily-trip threshold that defines treatment and the radius of the spillover buffer, and to the period over which e-scooter activity is measured.

All sensitivity models use the same specification as the main text, so the values are directly comparable to Tables~2--6 of the main text. Table~\ref{tab:definition} reports the results for the Intermediate Region.

\textbf{Threshold sweep.} We re-estimated the Bus+Metro and intermodal models under six alternative treatment thresholds: 1, 3, 5, 8, 10, and 15 trips per day. The Bus+Metro estimates range from $-1.2\%$ to $+0.5\%$ and are never statistically significant. The intermodal estimates are negative under all thresholds ($-2.6\%$ to $-6.8\%$), are statistically significant at the 5\% level for thresholds of 5 trips or more, and tend to be larger in absolute value under higher thresholds.

\textbf{Activity window.} Measuring e-scooter activity over all trips from March 1, 2019, instead of the study week, yields 49 treated zones for generation and 47 for attraction; the Bus+Metro estimates are $-0.3\%$ and $-0.5\%$ and the intermodal estimates $-2.7\%$ ($p=0.12$) and $-4.8\%$ ($p=0.051$).

\textbf{Buffer sweep.} We re-estimated under five alternative spillover-buffer radii: 0, 500, 1\,000, 1\,440, and 2\,000\,m. The exclusion set is derived with the same procedure as the main analysis (a buffer around the e-scooter trajectories), varying only the radius; the $1\,440$\,m row therefore reproduces the main estimates and their confidence intervals. The estimates are stable across the range: Bus+Metro $-0.7\%$ to $-1.1\%$ and intermodal $-4.7\%$ to $-5.7\%$.

\begin{table}[H]
\centering
\singlespacing
\scriptsize
\setlength{\tabcolsep}{3.5pt}
\begin{tabular}{@{}>{\raggedright\arraybackslash}p{3.1cm}cccc@{}}
\toprule
 & \multicolumn{2}{c}{Bus+Metro} & \multicolumn{2}{c}{Intermodal} \\
\cmidrule(lr){2-3}\cmidrule(lr){4-5}
 & Generation & Attraction & Generation & Attraction \\
\midrule
\multicolumn{5}{@{}l}{\emph{Daily e-scooter trip threshold}} \\
Threshold: $\geq$1 trips per day & \makecell{$+0.5$ [$-3.1$, $+4.2$] \\ $p$ = 0.849; 53/361} & \makecell{$-1.2$ [$-5.4$, $+3.1$] \\ $p$ = 0.691; 52/361} & \makecell{$-2.6$ [$-5.8$, $+0.8$] \\ $p$ = 0.150; 53/359} & \makecell{$-5.9$ [$-10.3$, $-1.2$] \\ $p$ = 0.017; 52/360} \\
Threshold: $\geq$3 trips per day & \makecell{$-0.9$ [$-3.3$, $+1.6$] \\ $p$ = 0.589; 42/361} & \makecell{$+0.2$ [$-4.2$, $+4.8$] \\ $p$ = 0.939; 42/361} & \makecell{$-3.2$ [$-6.5$, $+0.2$] \\ $p$ = 0.083; 42/359} & \makecell{$-4.6$ [$-9.1$, $+0.1$] \\ $p$ = 0.074; 42/360} \\
Threshold: $\geq$5 trips per day (main) & \makecell{$-0.9$ [$-3.3$, $+1.6$] \\ $p$ = 0.596; 38/361} & \makecell{$-1.1$ [$-3.8$, $+1.6$] \\ $p$ = 0.523; 38/361} & \makecell{$-4.7$ [$-7.4$, $-1.8$] \\ $p$ = 0.004; 38/359} & \makecell{$-5.7$ [$-9.5$, $-1.8$] \\ $p$ = 0.010; 38/360} \\
Threshold: $\geq$8 trips per day & \makecell{$-0.3$ [$-2.8$, $+2.3$] \\ $p$ = 0.874; 33/361} & \makecell{$-0.7$ [$-3.4$, $+2.2$] \\ $p$ = 0.719; 30/361} & \makecell{$-5.2$ [$-7.9$, $-2.5$] \\ $p<$0.001; 33/359} & \makecell{$-6.8$ [$-10.0$, $-3.6$] \\ $p$ = 0.002; 30/360} \\
Threshold: $\geq$10 trips per day & \makecell{$-0.1$ [$-2.6$, $+2.5$] \\ $p$ = 0.957; 32/361} & \makecell{$-0.7$ [$-3.6$, $+2.2$] \\ $p$ = 0.710; 26/361} & \makecell{$-5.2$ [$-7.9$, $-2.4$] \\ $p$ = 0.003; 32/359} & \makecell{$-6.7$ [$-9.8$, $-3.4$] \\ $p$ = 0.001; 26/360} \\
Threshold: $\geq$15 trips per day & \makecell{$+0.5$ [$-2.1$, $+3.2$] \\ $p$ = 0.768; 24/361} & \makecell{$-0.3$ [$-3.3$, $+2.8$] \\ $p$ = 0.879; 22/361} & \makecell{$-5.8$ [$-8.5$, $-3.0$] \\ $p$ = 0.003; 24/359} & \makecell{$-6.7$ [$-10.1$, $-3.3$] \\ $p$ = 0.003; 22/360} \\
\addlinespace
\multicolumn{5}{@{}l}{\emph{Period used to measure e-scooter activity}} \\
E-scooter trips from March 1, 2019 & \makecell{$-0.3$ [$-3.5$, $+3.0$] \\ $p$ = 0.905; 49/361} & \makecell{$-0.5$ [$-4.6$, $+3.8$] \\ $p$ = 0.872; 47/361} & \makecell{$-2.7$ [$-5.9$, $+0.5$] \\ $p$ = 0.124; 49/359} & \makecell{$-4.8$ [$-9.1$, $-0.3$] \\ $p$ = 0.051; 47/360} \\
\addlinespace
\multicolumn{5}{@{}l}{\emph{Radius of the exclusion buffer}} \\
Exclusion buffer: 0 m & \makecell{$-0.7$ [$-3.1$, $+1.8$] \\ $p$ = 0.674; 38/388} & \makecell{$-0.9$ [$-3.5$, $+1.9$] \\ $p$ = 0.627; 38/388} & \makecell{$-4.8$ [$-7.5$, $-2.0$] \\ $p$ = 0.002; 38/386} & \makecell{$-5.7$ [$-9.4$, $-1.9$] \\ $p$ = 0.007; 38/387} \\
Exclusion buffer: 500 m & \makecell{$-0.7$ [$-3.1$, $+1.8$] \\ $p$ = 0.664; 38/378} & \makecell{$-0.9$ [$-3.5$, $+1.9$] \\ $p$ = 0.626; 38/378} & \makecell{$-4.7$ [$-7.5$, $-1.9$] \\ $p$ = 0.005; 38/376} & \makecell{$-5.7$ [$-9.4$, $-1.9$] \\ $p$ = 0.009; 38/377} \\
Exclusion buffer: 1\,000 m & \makecell{$-0.8$ [$-3.2$, $+1.7$] \\ $p$ = 0.650; 38/371} & \makecell{$-1.0$ [$-3.6$, $+1.8$] \\ $p$ = 0.593; 38/371} & \makecell{$-4.7$ [$-7.4$, $-1.9$] \\ $p$ = 0.007; 38/369} & \makecell{$-5.7$ [$-9.4$, $-1.8$] \\ $p$ = 0.009; 38/370} \\
Exclusion buffer: 1\,440 m (main) & \makecell{$-0.9$ [$-3.3$, $+1.6$] \\ $p$ = 0.596; 38/361} & \makecell{$-1.1$ [$-3.8$, $+1.6$] \\ $p$ = 0.523; 38/361} & \makecell{$-4.7$ [$-7.4$, $-1.8$] \\ $p$ = 0.004; 38/359} & \makecell{$-5.7$ [$-9.5$, $-1.8$] \\ $p$ = 0.010; 38/360} \\
Exclusion buffer: 2\,000 m & \makecell{$-0.9$ [$-3.3$, $+1.6$] \\ $p$ = 0.607; 38/355} & \makecell{$-0.9$ [$-3.6$, $+1.8$] \\ $p$ = 0.606; 38/355} & \makecell{$-4.7$ [$-7.4$, $-1.8$] \\ $p$ = 0.004; 38/353} & \makecell{$-5.5$ [$-9.3$, $-1.6$] \\ $p$ = 0.009; 38/354} \\
\bottomrule
\end{tabular}
\caption{Sensitivity of Treatment $\times$ Post estimates to the daily-trip threshold used to define a treated zone, to the period used to measure e-scooter activity, and to the spillover-buffer radius, Intermediate Region. Cells as in Table~\ref{tab:periods}.}
\label{tab:definition}
\end{table}

\section{Intensity of E-scooter Use}
\label{sec:intensity}

Table~\ref{tab:intensity} relates the estimates to the intensity of e-scooter use in the treated zones of the Intermediate Region. The treated zones are divided into terciles of e-scooter trips per working day (arrivals for the generation model, departures for the attraction model), and each tercile is compared with the control zones. Intensity is the average over the working days of the study week and can therefore be below the treatment threshold, which a zone reaches when at least one of its zone-day observations has 5 trips or more. The last row of each block replaces the treatment indicator with a continuous measure of intensity, the logarithm of one plus the number of e-scooter trips per day, which equals zero in zones without e-scooter trips; for this model the randomization inference permutes intensity among the zones of the region.

\begin{table}[H]
\centering
\singlespacing
\scriptsize
\setlength{\tabcolsep}{3.5pt}
\begin{tabular}{@{}llrcc@{}}
\toprule
 & Treated zones & \makecell{E-scooter trips\\per day (range)} & Bus+Metro & Intermodal \\
\midrule
\multicolumn{5}{@{}l}{\emph{Generation}} \\
Low tercile & 14 & 0.4--3.4 & \makecell{$-4.7$ [$-7.4$, $-2.0$] \\ $p$ = 0.013; 14/337} & \makecell{$-1.3$ [$-6.2$, $+3.9$] \\ $p$ = 0.707; 14/335} \\
Middle tercile & 11 & 4.4--13.8 & \makecell{$-4.3$ [$-5.7$, $-2.9$] \\ $p$ = 0.001; 11/334} & \makecell{$-3.1$ [$-5.6$, $-0.5$] \\ $p$ = 0.064; 11/332} \\
High tercile & 13 & 14.2--97.4 & \makecell{$+2.6$ [$-0.0$, $+5.2$] \\ $p$ = 0.119; 13/336} & \makecell{$-7.8$ [$-10.7$, $-4.8$] \\ $p$ = 0.006; 13/334} \\
Continuous (per log unit) & 53 & -- & \makecell{$+0.1$ [$-0.7$, $+0.9$] \\ $p$ = 0.872; 53/361} & \makecell{$-1.7$ [$-2.5$, $-0.9$] \\ $p$ = 0.001; 53/359} \\
\addlinespace
\multicolumn{5}{@{}l}{\emph{Attraction}} \\
Low tercile & 13 & 0.2--2.6 & \makecell{$-4.2$ [$-8.5$, $+0.3$] \\ $p$ = 0.096; 13/336} & \makecell{$+4.5$ [$-10.4$, $+21.9$] \\ $p$ = 0.653; 13/335} \\
Middle tercile & 12 & 3.2--11.6 & \makecell{$-5.5$ [$-9.1$, $-1.8$] \\ $p$ = 0.014; 12/335} & \makecell{$-9.2$ [$-15.6$, $-2.2$] \\ $p$ = 0.044; 12/334} \\
High tercile & 13 & 11.8--115.8 & \makecell{$+1.6$ [$-1.0$, $+4.4$] \\ $p$ = 0.311; 13/336} & \makecell{$-5.9$ [$-8.9$, $-2.9$] \\ $p$ = 0.012; 13/335} \\
Continuous (per log unit) & 51 & -- & \makecell{$-0.1$ [$-0.9$, $+0.8$] \\ $p$ = 0.894; 51/361} & \makecell{$-2.1$ [$-3.1$, $-1.0$] \\ $p$ = 0.001; 51/360} \\
\bottomrule
\end{tabular}
\caption{Estimates by intensity of e-scooter use, Intermediate Region. Cells as in Table~\ref{tab:periods}; for the continuous measure, the number of zones with e-scooter trips is shown in place of the number of treated zones.}
\label{tab:intensity}
\end{table}

\section{Inferred Destinations}
\label{sec:destinations}

The smart card records boardings only. The trip tables published by DTPM contain, for each trip, the alighting point of every stage inferred by ADATRAP from the subsequent boardings of the same card. Trips whose last stage has no inferred alighting receive no expansion factor. The tables for 2018 and 2019 were produced with the same procedure. We do not have an independent source to validate the accuracy of the inferred alightings at the zone level.

Table~\ref{tab:destinations} reports the share of trips without an inferred alighting by the zone where the trip starts, in treated and control zones and in each year, and the difference-in-differences of that share, computed as the mean difference between treated and control zones in the change of the zone-level share, with a randomization-inference $p$-value. In the Intermediate Region the share is about one quarter in both groups and both years, and fell by 0.4 percentage points in treated zones and by 1.1 points in control zones.

\begin{table}[H]
\centering
\singlespacing
\scriptsize
\setlength{\tabcolsep}{3.5pt}
\begin{tabular}{@{}lrrrrrrr@{}}
\toprule
 & & \multicolumn{2}{c}{Treated zones (\%)} & \multicolumn{2}{c}{Control zones (\%)} & & \\
\cmidrule(lr){3-4}\cmidrule(lr){5-6}
Region & Treated/zones & 2018 & 2019 & 2018 & 2019 & DiD (pp) & $p$ (RI) \\
\midrule
Central & 2/58 & 34.2 & 34.8 & 23.9 & 22.8 & $+1.74$ & 0.249 \\
Intermediate & 38/361 & 22.9 & 22.5 & 25.9 & 24.8 & $+0.61$ & $<$0.001 \\
Peripheral & 2/320 & 28.9 & 29.1 & 29.0 & 27.9 & $+0.58$ & 0.436 \\
\bottomrule
\end{tabular}
\caption{Share of trips without an inferred alighting (\%), by region, group of zones (by boarding zone) and year, for the working days of the study weeks. Treated zones are those of the generation model. DiD: difference-in-differences of the zone-level share, in percentage points.}
\label{tab:destinations}
\end{table}

\section{Availability-Based Treatment Definition}
\label{sec:availability}

The main analysis defines treated zones by observed e-scooter use. To separate the effect of the service from the selection of the zones where it was used, we re-estimated all models with a definition based on service availability (intention to treat). The service started in the municipality of Las Condes, and from February 2019 it extended to parts of the neighboring municipalities of Vitacura and La Reina. In the availability-based definition, all zones of Las Condes are treated, whether or not e-scooters were used in them (27 zones in the Intermediate Region, 4 in the Peripheral Region and 1 in the Central Region); zones outside Las Condes that reach the use threshold are excluded from the control group, and the spillover buffer is maintained. As a variant, all zones of Las Condes, Vitacura and La Reina are treated (56 zones in the Intermediate Region).

Table~\ref{tab:availability} reports the results for the Intermediate Region next to those of the main analysis. The conclusions are the same. With the Las Condes definition, the Bus+Metro effect is not statistically significant ($+0.5\%$ for generation and $-0.8\%$ for attraction), and intermodal trips decline ($-5.0\%$, $p=0.028$, and $-5.8\%$, $p=0.007$). With the wider area, the intermodal estimates are $-3.0\%$ ($p=0.096$) and $-5.4\%$ ($p=0.039$). The placebo test between April and May 2018 for the availability-based treated zones is close to zero for the Bus+Metro and intermodal models (last column).

\begin{table}[H]
\centering
\singlespacing
\scriptsize
\setlength{\tabcolsep}{3.5pt}
\begin{tabular}{@{}lcccc@{}}
\toprule
 & \makecell{Use-based\\(main analysis)} & \makecell{Availability:\\Las Condes} & \makecell{Availability: Las Condes,\\Vitacura and La Reina} & \makecell{Availability: Las Condes;\\placebo April $\rightarrow$ May 2018} \\
\midrule
\multicolumn{5}{@{}l}{\emph{Generation}} \\
Bus+Metro & \makecell{$-0.9$ [$-3.3$, $+1.6$] \\ $p$ = 0.596; 38/361} & \makecell{$+0.5$ [$-2.2$, $+3.3$] \\ $p$ = 0.779; 27/347} & \makecell{$-0.6$ [$-3.6$, $+2.6$] \\ $p$ = 0.793; 56/362} & \makecell{$-0.9$ [$-1.9$, $+0.2$] \\ $p$ = 0.214; 27/347} \\
Bus & \makecell{$-2.9$ [$-4.9$, $-0.8$] \\ $p$ = 0.009; 38/361} & \makecell{$-2.8$ [$-5.5$, $-0.1$] \\ $p$ = 0.078; 27/347} & \makecell{$-1.5$ [$-4.2$, $+1.2$] \\ $p$ = 0.346; 56/362} & \makecell{$-2.3$ [$-3.5$, $-1.0$] \\ $p$ = 0.001; 27/347} \\
Metro & \makecell{$+0.5$ [$-2.6$, $+3.7$] \\ $p$ = 0.865; 8/61} & \makecell{$+0.5$ [$-2.6$, $+3.7$] \\ $p$ = 0.868; 8/61} & \makecell{$-16.3$ [$-42.5$, $+21.9$] \\ $p$ = 0.401; 11/62} & \makecell{$-0.3$ [$-1.7$, $+1.1$] \\ $p$ = 0.748; 8/58} \\
Intermodal & \makecell{$-4.7$ [$-7.4$, $-1.8$] \\ $p$ = 0.004; 38/359} & \makecell{$-5.0$ [$-8.6$, $-1.1$] \\ $p$ = 0.028; 27/345} & \makecell{$-3.0$ [$-6.2$, $+0.3$] \\ $p$ = 0.096; 56/360} & \makecell{$+0.8$ [$-1.0$, $+2.6$] \\ $p$ = 0.476; 27/346} \\
\addlinespace
\multicolumn{5}{@{}l}{\emph{Attraction}} \\
Bus+Metro & \makecell{$-1.1$ [$-3.8$, $+1.6$] \\ $p$ = 0.523; 38/361} & \makecell{$-0.8$ [$-3.7$, $+2.1$] \\ $p$ = 0.662; 27/347} & \makecell{$-2.6$ [$-6.4$, $+1.2$] \\ $p$ = 0.286; 56/362} & \makecell{$-1.3$ [$-3.0$, $+0.5$] \\ $p$ = 0.192; 27/347} \\
Bus & \makecell{$-3.5$ [$-6.5$, $-0.4$] \\ $p$ = 0.032; 38/361} & \makecell{$-4.2$ [$-7.5$, $-0.8$] \\ $p$ = 0.020; 27/347} & \makecell{$-2.7$ [$-5.5$, $+0.3$] \\ $p$ = 0.085; 56/362} & \makecell{$-2.2$ [$-4.2$, $-0.2$] \\ $p$ = 0.027; 27/347} \\
Metro & \makecell{$-10.8$ [$-36.0$, $+24.2$] \\ $p$ = 0.559; 20/105} & \makecell{$-11.7$ [$-32.6$, $+15.6$] \\ $p$ = 0.437; 15/99} & \makecell{$-23.0$ [$-48.0$, $+13.9$] \\ $p$ = 0.238; 27/107} & \makecell{$+2.0$ [$-17.3$, $+25.9$] \\ $p$ = 0.857; 15/89} \\
Intermodal & \makecell{$-5.7$ [$-9.5$, $-1.8$] \\ $p$ = 0.010; 38/360} & \makecell{$-5.8$ [$-9.1$, $-2.4$] \\ $p$ = 0.007; 27/346} & \makecell{$-5.4$ [$-10.1$, $-0.4$] \\ $p$ = 0.039; 56/361} & \makecell{$-0.5$ [$-3.3$, $+2.5$] \\ $p$ = 0.799; 27/346} \\
\bottomrule
\end{tabular}
\caption{Use-based and availability-based definitions of treated zones, Intermediate Region. Cells as in Table~\ref{tab:periods}. In the availability-based definitions all zones of the indicated municipalities are treated, regardless of e-scooter use.}
\label{tab:availability}
\end{table}
\end{document}